\documentclass[11pt]{article}
\usepackage{amsmath}
\usepackage{mathrsfs}
\usepackage{bm}
\usepackage{hyperref}
\usepackage[at]{easylist}
\usepackage{here}
\usepackage{cite}
\usepackage{amssymb}
\usepackage{mhchem}
\usepackage{physics}
\usepackage{authblk}
\usepackage{comment}
\usepackage{indentfirst}
\usepackage{multirow}
\usepackage[subrefformat=parens]{subcaption}
\usepackage[top=30truemm,bottom=30truemm,left=20truemm,right=20truemm]{geometry}
\usepackage{color}

\usepackage{makecell}

\makeatletter
\@addtoreset{equation}{section}

\makeatother

\newcommand{\Lop}{\mathbb{L}}
\newcommand{\LopD}{\mathbb{L}_{D}}
\newcommand{\Lopk}{\mathbb{L}_{k}}
\newcommand{\LopDk}{\mathbb{L}_{Dk}}

\newcommand{\Sh}{\mathcal{S}}  
\newcommand{\QA}{\mathcal{Q}} 
\newcommand{\Lps}{\ell}

\newcommand{\pt}[2]{\partial_{#2} #1}
\newcommand{\ptt}[2]{\partial_{#2}^{2} #1}

\title{Anisotropic separate universe:\\ Long-wavelength perturbations and conserved quantities
}
\author{Daiki Saito$^{1}$\footnote{dsaito@ewha.ac.kr}}
\author{Atsushi Naruko$^{2,3,4,5}$\footnote{naruko@yukawa.kyoto-u.ac.jp}}

\affil{$^{1}$Department of Science Education, Ewha Womans University, Seoul 03760, Korea}
\affil{$^{2}$
National Institute of Technology, Fukui College, 
Geshicho, Sabae, Fukui 916-8507, Japan}
\affil{$^{3}$Center for Gravitational Physics and Quantum Information, \\
Yukawa Institute for Theoretical Physics, Kyoto University, Kyoto 606-8502, Japan}
\affil{$^{4}$Asia Pacific Center for Theoretical Physics, Pohang 37673, Korea}
\affil{$^{5}$ 
Faculty of Engineering Science, Yokohama National University, Yokohama 240-8501, Japan}

\begin{document}

\maketitle

\begin{abstract}
We study the superhorizon evolution of linear perturbations in an anisotropic Bianchi type I universe with a scalar field coupled to a U(1) gauge field. 
The analysis employs the spatial gradient expansion in the uniform-\(\mathcal{N}\) gauge, in which the local number of $e$-folds is unperturbed. 
In this gauge, the complete set of leading-order long-wavelength solutions, including both even- and odd-parity perturbations, can be constructed from variations of homogeneous Bianchi type I solutions, subject to the momentum and Gauss constraints.
The conserved Wronskian found in isotropic multi-field inflation is generalized to the anisotropic system. Together with the shear and electric-flux integration constants, it allows the curvature perturbation on uniform-\(\mathcal{N}\) hypersurfaces to be expressed in terms of background variations and time integrals. 
The conserved comoving curvature perturbations are then obtained by adding this intrinsic curvature contribution to the familiar $\delta N$ contribution arising from the change of time slicing.
We also derive a relation between the curvature perturbation and a metric perturbation that reduces to the plus-polarized tensor mode in the isotropic limit, and discuss a nonlinear extension of the formalism.
\end{abstract}

\tableofcontents

\section{Introduction}
\label{Intro}
Cosmic inflation~\cite{Brout:1977ix,Guth:1980zm,Sato:1980yn,Starobinsky:1980te,Linde:1981mu} provides a compelling explanation for the origin of primordial cosmological perturbations. 
While most inflationary models assume a homogeneous and isotropic background, models involving vector fields can support anisotropic inflationary solutions. 
Such models can generate statistical anisotropy in the cosmic microwave background (CMB)~\cite{Ackerman:2007nb,Dimopoulos:2008yv,Pitrou:2008gk,Yokoyama:2008xw,Kanno:2008gn,Gumrukcuoglu:2010yc,Dulaney:2010sq,Watanabe:2010fh,Kanno:2010nr,Soda:2012zm,Bartolo:2012sd,Naruko:2014bxa,Fujita:2017lfu} (see Refs.~\cite{Kim:2013gka,Planck:2019evm} for observational constraints) and large-scale structure~\cite{pullen2010non,Sugiyama:2017ggb}, and provide counterexamples to the cosmic no-hair conjecture~\cite{Wald:1983ky}, as shown in Refs.~\cite{Golovnev:2008cf,Kanno:2008gn,Watanabe:2009ct,Gong:2019hwj} (see Refs.~\cite{Kim:2013gka,Planck:2019evm} for observational constraints) and large-scale structure~\cite{pullen2010non,Sugiyama:2017ggb}, and provide counterexamples to the cosmic no-hair conjecture~\cite{Wald:1983ky}, as shown in Refs.~\cite{Golovnev:2008cf,Kanno:2008gn,Watanabe:2009ct,Gong:2019hwj}. 
A prototypical example is a scalar field coupled to a U(1) gauge field, which admits homogeneous but anisotropic Bianchi type I inflationary solutions~\cite{Watanabe:2009ct}. In such backgrounds, the reduced rotational symmetry leads to mixing among perturbations that would be classified as scalar, vector, and tensor modes in isotropic cosmology~\cite{Gumrukcuoglu:2007bx,Pereira:2007yy}. Understanding their superhorizon evolution is therefore important for studying primordial perturbations in anisotropic inflation.

To study the superhorizon evolution of cosmological perturbations, the spatial gradient expansion provides a systematic description of long-wavelength perturbations~\cite{Bardeen:1980kt,Starobinsky:1985ibc,Salopek:1990jq,Deruelle:1994iz,Nambu:1994hu,Kodama:1997qw,Shibata:1999zs,Wands:2000dp,Rigopoulos:2003ak,Lyth:2004gb}. 
For isotropic Friedmann--Lema\^itre--Robertson--Walker (FLRW) backgrounds, Ref.~\cite{Sasaki:1998ug} established a one-to-one correspondence between long-wavelength perturbations and infinitesimal variations of homogeneous background solutions in multi-scalar-field models. 
The same work also derived a conserved Wronskian that determines the integrated contribution to the curvature perturbation on uniform-$\mathcal{N}$ hypersurfaces. 
These results provide a concrete realization of the separate-universe picture underlying the $\delta N$ formalism~\cite{Sasaki:1995aw,Tanaka:2006zp,Naruko:2012fe,Sugiyama:2012tj,Artigas:2024ajh}.
Recent work has also highlighted the role of anisotropic perturbations in the momentum constraint for describing non-adiabatic modes beyond slow roll~\cite{Artigas:2025nbm}.

The extension to anisotropic cosmologies requires additional care because of the presence of shear and vector fields. 
The $\delta N$ formalism has been applied to specific models of anisotropic inflation~\cite{Abolhasani:2013zya}. 
More generally, Ref.~\cite{Tanaka:2021dww} established an anisotropic separate-universe framework, showing that the long-wavelength evolution is governed by homogeneous equations, provided that the momentum constraints and Gauss's law are satisfied. 
Based on this framework, the generalized $\delta N$ ($g\delta N$) formalism~\cite{Tanaka:2023gul,Tanaka:2024mzw} enables the evaluation of large-scale curvature and tensor perturbations from homogeneous background evolution. 
An alternative approach, known as the $\delta M$ formalism, has also been developed for anisotropic inflation~\cite{Talebian-Ashkezari:2016llx}.

These works establish the general framework describing superhorizon evolution in anisotropic inflation.
In isotropic backgrounds, the traceless part of the spatial metric affects the curvature perturbation only through decaying modes, controlled by the conserved Wronskian of Ref.~\cite{Sasaki:1998ug}, while its time-independent part can be removed by a spatial coordinate transformation; nonlinear treatments such as the $\delta N$ formalism therefore neglect it by assuming that the shear has decayed~\cite{Salopek:1990jq,Lyth:2004gb}.
In anisotropic backgrounds, the shear is sourced by the anisotropic stress of the vector field and does not decay, and the reduced rotational symmetry prevents even a time-independent traceless metric from being removed, so that the traceless metric enters the curvature of uniform-$\mathcal N$ hypersurfaces at the same order in spatial gradients as the conformal factor.
Although formally contained in the framework of Refs.~\cite{Tanaka:2021dww,Tanaka:2023gul,Tanaka:2024mzw}, this contribution has not been evaluated analytically, even at linear order.
In particular, the explicit mode structure and the background--perturbation correspondence at the level of individual perturbation equations remain to be clarified for the scalar--U(1) system, and the constraint equations, the conserved Wronskian, and the intrinsic curvature on uniform-$\mathcal N$ hypersurfaces have not been systematically connected through analytic expressions in terms of background solutions.

In this paper, we extend the analysis of Ref.~\cite{Sasaki:1998ug} to anisotropic Bianchi type I backgrounds sourced by a scalar field and a U(1) gauge field. 
Using the spatial gradient expansion in the uniform-$\mathcal{N}$ gauge, in which the local number of $e$-folds is unperturbed, the complete set of leading-order linear perturbations is constructed from homogeneous background solutions, subject to the momentum and Gauss constraints. 
The main purpose of this paper is to evaluate the contribution of the traceless metric to the curvature analytically at linear order, to identify the conserved quantities that control its decaying part, and to make explicit the non-decaying and constant parts that have no isotropic counterpart.

Our main results are summarized as follows:

\begin{itemize}

\item
We construct the complete set of leading-order long-wavelength linear perturbations for generic wave vectors from variations of homogeneous Bianchi type I solutions. 
While part of the even-parity sector corresponds directly to variations of the axisymmetric background, the remaining even-parity perturbations and the odd-parity sector can be understood as non-axisymmetric deformations of general Bianchi type I solutions. 
The momentum and Gauss constraints relate their integration constants. 
The solutions contain six physical parameters in the even-parity sector and four in the odd-parity sector, corresponding to five propagating degrees of freedom. 
Among the even-parity solutions, the time-translation mode is pure gauge, while three other directions arise from global coordinate and U(1) transformations that become nontrivial for spatially varying perturbations.

\item
We derive a conserved quantity $W$ that generalizes the Sasaki--Tanaka Wronskian~\cite{Sasaki:1998ug}. 
Together with the shear and electric-flux integration constants and the dynamical metric shear $D$, it allows the intrinsic curvature on uniform-$\mathcal N$ hypersurfaces to be evaluated explicitly in terms of background variations and time integrals. 
Together with the usual $\delta N$ contribution, this intrinsic curvature determines the two directional comoving curvature perturbations during the anisotropic stage.

\item
The dilatation mode, corresponding to the adiabatic growing mode in the isotropic limit, carries nonzero variations of the shear and electric-flux charges even though its Wronskian vanishes. 
Consequently, the comoving curvature perturbation can evolve even for this mode alone, within the residual spatial gauge fixed in Sec.~\ref{SS}. 
For attractor-like backgrounds with nearly constant $H$ and small shear, the memory integrals approach constants within a few $e$-folds, but the curvature can continue to evolve through the changing physical direction of the wave vector. 
The anisotropic rescaling of spatial coordinates also produces a direction-dependent constant contribution to the curvature, absent from the scalar-field analysis of Ref.~\cite{Sasaki:1998ug}.

\item
We derive an explicit relation between the curvature perturbation and a metric perturbation that reduces to the plus-polarized tensor mode in the isotropic limit. 
This relation holds during the anisotropic stage and complements previous perturbative calculations of curvature--tensor cross-correlations~\cite{Gumrukcuoglu:2010yc,Watanabe:2010fh,Chen:2014eua} and the generalized $\delta N$ approach~\cite{Tanaka:2023gul,Tanaka:2024mzw}. 

\end{itemize}

A nonlinear extension of the background--perturbation correspondence is also discussed in the growing-mode sector.

This paper is organized as follows. In Sec.~\ref{ABGPT}, we introduce the background and the perturbation variables. 
In Sec.~\ref{LWP}, we construct the long-wavelength solutions and derive the conserved Wronskian. 
In Sec.~\ref{R and G}, we evaluate the spatial curvature and its relation to tensor metric perturbations. Sec.~\ref{Con} contains our conclusions. 
The appendices provide further details, including the odd-parity solutions and a nonlinear extension. Throughout this paper, we use units in which $c=8\pi G=1$.

\section{Anisotropic background and linear perturbation theory}
\label{ABGPT}

In this section, we first review the homogeneous and anisotropic background spacetime in Sec.~\ref{BG}, and describe the parameter space of the background solutions in Sec.~\ref{BGparam}.
Then, in Secs.~\ref{Dec} and ~\ref{SPG}, we develop the decomposition of linear perturbations around it, fixing the gauge for the rest of the paper.
    
\subsection{Background equations}
\label{BG}

In this paper, we consider the following action introduced in Ref.~\cite{Watanabe:2009ct}
\begin{align}
    S=\int d^4x\sqrt{-g}\qty[\frac{1}{2}R-\frac{1}{2}g^{\mu\nu}\partial_{\mu}\phi\partial_{\nu}\phi-U(\phi)-\frac{1}{4}I^2(\phi)F_{\mu\nu}F^{\mu\nu}],\label{eq:S}
    \end{align}
where $\phi$ is a real scalar field and $U(\phi)$ and $I(\phi)$ are functions of the scalar field.
$F_{\mu\nu}$ is the field strength of a U(1) gauge field $A_{\mu}$ coupled with the scalar field through $I(\phi)$.

We assume the spatial homogeneity of the background spacetime and that of the matter fields.

Using the U(1) gauge degree of freedom to remove the time component, and choosing the $x$-axis along the direction of the vector field, we take
$A_{\mu}=\bar{A}_{x}(t) \, \delta^{x}_{\mu}$.
We also assume that the background geometry is described by the Bianchi type I metric:
\begin{align}
    ds^2=-dt^2+e^{2\alpha(t)}\qty[e^{-4\beta(t)}dx^2+e^{2\beta(t)}\qty(e^{2\sqrt{3}\gamma(t)}dy^2+e^{-2\sqrt{3}\gamma(t)}dz^2)]  
=:\bar{g}_{\mu\nu}(t)dx^{\mu}dx^{\nu}. 
    \label{eq:B}
    \end{align}
Here, $e^{\alpha}$ is the mean scale factor, while $\beta$ and $\gamma$ parametrize the anisotropic expansion. The Hubble parameter $H$ and the $e$-folding number $N$ are defined as
\begin{align}
    & H(t) := \frac{\dd \alpha(t)}{\dd t} , \quad  
    N(t) :
    =\int^t H(t_*) dt_*
    = \int^t \frac{\dd \alpha(t_*)}{\dd t_*} dt_*. 
    \notag
    \end{align} 
In the following, we take $N$ as the time coordinate, choosing its origin such that $\alpha=N$.

We express the components of the energy-momentum tensor $T_{\mu\nu}$ as
\begin{align}
    &\rho = T_{\mu\nu}n^{\mu}n^{\nu}, \quad J_{i} = -T_{i \mu}n^{\mu}, \quad P = \frac{1}{3}\sum_{i=1}^{3}T^{i}_{\ i}, \quad \pi^{i}_{\, j} = T^{i}_{\ j} - P\delta^{i}_{j}, \quad (i,j=x,y,z)
    \notag
 \end{align}
where $\rho$ is the energy density, $J_{i}$ is the momentum current, $P$ is the pressure, and $\pi^{i}_{\, j}$ denotes the anisotropic stress.
$n^{\mu} =$ $H\delta^{\mu}_{N}$ is the unit normal to the constant-time hypersurface.
These definitions are also used for the perturbed quantities in Sec.~\ref{PE} and App.~\ref{PEspacetime}.
The concrete form of the components reads
\begin{align}
    &\rho=\frac{1}{2}H^2 \phi'^2+U(\phi)+\frac{1}{2}\mathcal{V}, \quad J_{i}=0, \quad 
    P=\frac{1}{2}H^2  \phi'^2-U(\phi)+\frac{1}{6}\mathcal{V}, 
    \notag 
\end{align}
  and
\begin{align}
 \pi^{x}_{\ x} = -\frac{2}{3}\mathcal{V}, \quad 
 \pi^{a}_{\ b} = \frac{1}{3}\mathcal{V}\delta^{a}_{b}, \quad (a,b=y,z)
    \notag 
    \end{align}
 where
 \begin{align}
   \mathcal{V}&:=H^2I^2\bar{A}'^{2}_{x}\,e^{-2\alpha+4\beta} \, \label{eq:nu}
 \end{align}
is twice the energy density of the vector field, and the prime denotes the derivative with respect to $N$.

The equation of motion for the scalar field is given by
\begin{align}
&H\frac{d}{dN}\qty(H\phi')+3H^2\phi'+\pt{U}{\phi}=\frac{\partial_{\phi}I}{I}\mathcal{V}.
    \label{eq:ABGphi}
\end{align}
Since there is no charged matter field, the equations of motion for the vector field are written as
\begin{align}
&\frac{1}{\sqrt{-g}}\partial_{\nu}\qty(\sqrt{-g}I^2F^{\nu\mu})=0.
    \label{eq:BGMax} 
\end{align}
Here, $g$ denotes the determinant of the metric.
Only the $x$-component is non-zero, which can be integrated to give
\begin{align}
    & \bar{A}'_{x} = \frac{d_{A}}{HI^2e^{\alpha+4\beta}}, \label{eq:usof}
\end{align}
with $d_{A}$ being an integration constant.
In the perturbation formulas below that involve $1/d_A$ or $1/\bar A_x^{\prime}$, we assume a nonvanishing background electric flux, $d_A\neq0$.
For comparison with the isotropic analysis of Ref.~\cite{Sasaki:1998ug}, we also refer to the model in which the gauge field is absent from the outset, so that $d_{A}$ and the quantities derived from it are not defined, as the \emph{scalar-only case}.
It should be distinguished from the isotropic limit $d_{A}\rightarrow0$ taken in the formulas derived for $d_{A}\neq0$, which is defined in Sec.~\ref{W1}.
Using this solution, $\mathcal{V}$ is expressed as 
\begin{align}
   \mathcal{V}&=\frac{d_{A}^2}{I^{2}e^{4(\alpha+\beta)}}
   =\frac{d_{A}H\bar{A}'_{x}}{e^{3\alpha}}. \label{eq:nu2}
 \end{align}

The Einstein equations are
\begin{align}
&- 3H^2\qty(1 - \beta'^2 - \gamma'^2) = -\rho, \label{eq:ABG00} \\
&2HH' +3H^2\qty(1 + \beta'^2 + \gamma'^2) = - P,\label{eq:ABGtr} \\
&H\frac{d}{dN}(H\beta')+3H^2\beta' = \frac{1}{3} \mathcal{V}, \label{eq:Abeta1eq} \\
&H\frac{d}{dN}(H\gamma')+3H^2\gamma' = 0.
\label{eq:Abeta2eq}
\end{align}
We note that the equations~\eqref{eq:ABGphi}, \eqref{eq:ABG00}, \eqref{eq:ABGtr}, \eqref{eq:Abeta1eq}, and \eqref{eq:Abeta2eq} are not independent, and only four of them are independent.

We can integrate Eqs.~\eqref{eq:Abeta1eq} and \eqref{eq:Abeta2eq} as
\begin{align}
    \beta' &= \frac{1}{He^{3\alpha}}\qty(d_{\beta}+\frac{1}{3}d_{A}\bar{A}_{x}), 
    \label{eq:Acbeta} \\
    \gamma'&=\frac{d_{\gamma}}{He^{3\alpha}},
\end{align}
where  $d_{\beta}$ and $d_{\gamma}$ are new constants of integration.
From Eq.~\eqref{eq:Acbeta}, we see that the vector field acts as a source that drives the growth of the anisotropy associated with $\beta$.
For later use, we define
\begin{align}
 &I_{\alpha}(N) := \int^N_{N_{0}} \frac{dN_*}{He^{3\alpha}}. \label{eq:Ia}
\end{align}
Here $N_{0}$ is a fixed reference time, which will be identified with the initial time of the superhorizon evolution in Sec.~\ref{sec:flat}.
We use the same lower limit $N_{0}$ for all the time integrals introduced below.
In contrast, as the Universe expands, $\gamma'$ decays as $e^{-3\alpha}/H$, since it has no corresponding source term.
Consequently, after a sufficient time, $\gamma$ approaches a constant and can be absorbed into the coordinates $y$ and $z$.
Hence, in the remainder of this paper, we set $\gamma=0$ and consider the spacetime
\begin{align}
\bar{g}_{\mu\nu}dx^{\mu}dx^{\nu} \simeq
    - \frac{dN^2}{H^2}+e^{2\alpha}\qty[e^{-4\beta}dx^2+e^{2\beta}\qty(dy^2+dz^2)]  \label{eq:B1}
    \end{align}
as the background.
Although we set $\gamma=0$ in the background, its variation is not lost at the level of perturbations.
The trace-free perturbation $f$ in the $(y,z)$-plane obeys the same equation as $\gamma$, as we show in Sec.~\ref{SS}.
Its integration constant $d_f$, introduced in Sec.~\ref{SS}, thus plays the role of the variation of $d_{\gamma}$.

\subsection{Parameter space of background solutions}
\label{BGparam}

After setting $\gamma=0$, the dynamical variables of the background are $\beta$, $\phi$ and $\bar{A}_{x}$, and $H$ is fixed by the Hamiltonian constraint~\eqref{eq:ABG00}.
We introduce the shorthand notation $\chi^{m}:=(\beta, \phi,\bar{A}_{x})$ for the set of dynamical fields.
Each component obeys a second-order differential equation in time, so the set of the solutions of $\chi^{m}$ is characterized by six integration constants corresponding to the initial values of $\beta, \beta', \phi, \phi', \bar{A}_{x}, \bar{A}'_{x}$.
One of them corresponds to translation of the global time coordinate $N$. 

Also, we see that a constant shift of $\alpha$ alone does not leave the background equations invariant, but they are invariant under the combined transformation
\begin{align}
x^i \rightarrow e^{-\bar{s}} x^i, \quad \alpha \rightarrow \alpha + \bar{s}, \quad \bar{A}_{x} \rightarrow e^{\bar{s}} \bar{A}_{x}, \quad d_{A} \rightarrow e^{2\bar{s}}  d_{A},  \quad d_{\beta} \rightarrow e^{3\bar{s}}  d_{\beta},
    \end{align}
where $\bar{s}$ is a constant.
This transformation is nothing but the dilatation.
In the following, we denote the corresponding variation by $\Delta_{s}$. 
In the isotropic case of Ref.~\cite{Sasaki:1998ug}, the corresponding freedom appears only as an integration constant of the curvature perturbation (see Sec.~\ref{directions}), and not as a parameter of the background solution.

Thus, the homogeneous solutions can be parametrized by six integration constants---one time-translation parameter and five constants $\lambda^{p}$ $(p=1,\dots,5)$---together with the dilatation parameter $\bar{s}$.
The latter corresponds to a rescaling of spatial coordinates in the strictly homogeneous limit. 
When the time translation is included, we write the six integration constants collectively as $\lambda^{J}=(N, \lambda^{p})$ $(J=1,\dots,6)$.
In the rest of this subsection, we give a concrete choice of $\lambda^{p}$ and discuss two more symmetries of the background.

The background admits one more rescaling symmetry, the volume-preserving
anisotropic rescaling $x\to e^{2\sigma}x$, $x^a\to e^{-\sigma}x^a$:
\begin{equation}
  \beta\to\beta+\sigma,\qquad
  \bar{A}_{x}\to e^{-2\sigma}\bar{A}_{x},\qquad
  d_{A}\to e^{2\sigma}d_{A},\qquad
  d_{\beta}\to d_{\beta},
  \label{eq:aniso_dil}
\end{equation}
with $\alpha$, $\mathcal{V}$ and $d_{A} \bar{A}_{x}$ being invariant.

For later use, let us consider the concrete form of the background parameter $\lambda^{p}$.
We assume that $\phi$ is a monotonic function of $N$ in the time interval considered, so that the parametrization below is a local chart of the solution space.
We may then use $\phi$ as a clock $\phi = \phi(N)$, and fix the time-translation freedom by evaluating the remaining data on the slice $\phi = \phi_0$.
Furthermore, instead of $\beta', \bar{A}'_{x}$, we may use the constants of motion $d_{A}$ and $d_{\beta}$ and adopt
\begin{align}
\lambda^{p} = \qty(\phi'_{0}, \beta_{0},  \bar{A}_{x0}, d_{A}, d_{\beta}),\label{eq:lambda}
    \end{align}
where the subscript $0$ denotes the value on the reference slice $\phi=\phi_{0}$, so that $(\phi'_{0}, \beta_{0}, \bar{A}_{x0})$ are data on this slice, while $(d_{A}, d_{\beta})$ are the conserved charges of Eqs.~\eqref{eq:usof} and \eqref{eq:Acbeta}.
The reference slice is fixed once and for all and need not coincide with the reference time $N_{0}$ introduced in Eq.~\eqref{eq:Ia}.
In what follows, we label the components of $\lambda^{p}$ by these names, e.g., $\partial_{\beta_{0}}:=\partial/\partial\beta_{0}$ at fixed values of the other parameters, and write the coefficients of a variation as $c^{p}\partial_{\lambda^{p}} = c^{\phi'_{0}}\partial_{\phi'_{0}} + c^{\beta_{0}}\partial_{\beta_{0}} + c^{\bar{A}_{x0}}\partial_{\bar{A}_{x0}} + c^{d_{A}}\partial_{d_{A}} + c^{d_{\beta}}\partial_{d_{\beta}}$.
Since the action depends on $\bar{A}_{x}$ only through $F_{\mu\nu}$, the origin of $\bar{A}_{x}$ is arbitrary. 
For the reference background we fix it by $\bar{A}_{x0}=0$, so that $d_{\beta}=He^{3\alpha}\beta'|_{\phi=\phi_{0}}$ is the shear charge measured on the reference slice, while the variation $c^{\bar{A}_{x0}}$ remains a free parameter.
We can show that this choice of $\lambda^p$ is well-defined as follows.
Given these $\lambda^p$, Eq.~\eqref{eq:nu2} fixes $\mathcal{V}$.
By solving Eq.~\eqref{eq:Acbeta} as $He^{3\alpha}\beta' = d_{\beta} + d_{A}\bar{A}_{x}/3$ and substituting it into Eq.~\eqref{eq:ABG00} with $\gamma = 0$, we obtain
\begin{align}
H^2\qty(3-\frac{\phi'^2}{2}) = U + \frac{1}{2}\mathcal{V} + 3\qty(d_{\beta}+\frac{1}{3}d_{A}\bar{A}_{x})^2e^{-6\alpha}. \label{eq:Hrecon}
    \end{align}
From this relation, we can determine $H$ algebraically.
Equation~\eqref{eq:ABG00} implies $\phi'^2<6$ whenever $U+\mathcal{V}/2>0$. Under this assumption, Eq.~\eqref{eq:Hrecon} determines $H$ locally on the expanding branch.
Together with the assumed monotonicity of $\phi$, this makes the map $\qty(\beta, \beta', \phi', \bar{A}_{x}, \bar{A}'_{x}) \rightarrow \qty(\phi'_{0}, \beta_{0}, \bar{A}_{x0}, d_{A}, d_{\beta})$ locally invertible, so the two sets label the same five-parameter family of background solutions locally.

Unlike $\Delta_s$, the anisotropic rescaling~\eqref{eq:aniso_dil} requires no additional parameter, as it lies within the span of the $\partial_{\lambda^p}$:
\begin{equation}
  \Delta_\sigma
  = \partial_{\beta_{0}}
  - 2\bar{A}_{x0}\,\partial_{\bar{A}_{x0}}
  + 2d_{A}\,\partial_{d_{A}}.
  \label{eq:Dsigma}
\end{equation}
With our convention $\bar{A}_{x0}=0$, the second term vanishes.

The difference between $\Delta_s$ and $\Delta_\sigma$ stems from our use of $\alpha=N$ as the time coordinate.
The normalization of $e^{\alpha}$ relative to the comoving coordinates is fixed by the choice of the time coordinate and is therefore not among the initial data $\lambda^{p}$.
Hence, the dilatation requires the separate parameter $\bar{s}$.
In contrast, the constant part of $\beta$ is part of the initial data ($\beta_{0}\in\lambda^{p}$), so that the anisotropic rescaling is already contained in the variations of $\lambda^{p}$.
Similarly, the background is invariant under the constant shift $\bar{A}_{x}\rightarrow\bar{A}_{x}+\bar{\theta}$ together with $d_{\beta}\rightarrow d_{\beta}-d_{A}\bar{\theta}/3$, as follows from Eq.~\eqref{eq:Acbeta}.
This direction,
\begin{equation}
\Delta_{\theta}:=\partial_{\bar{A}_{x0}}-\frac{d_{A}}{3}\partial_{d_{\beta}}, \label{eq:Dtheta}
\end{equation}
is also contained in the variations of $\lambda^{p}$ and corresponds to a residual U(1) gauge transformation with $\theta=\bar{\theta}x$, discussed in Sec.~\ref{SPG}.

Having established the structure of the background, we now turn to the linear perturbations around it.

\subsection{Decomposition of perturbation variables}
\label{Dec}

In this subsection, we discuss the decomposition of the perturbed variables according to the symmetry of the background geometry~\eqref{eq:B1}. We also count the number of degrees of freedom in each sector.

Around a homogeneous and isotropic background, the three-dimensional rotational symmetry allows the standard scalar-vector-tensor (SVT) decomposition, and the three sectors evolve independently.
However, in an anisotropic background~\eqref{eq:B1}, the full three-dimensional rotational symmetry is broken, leaving only a two-dimensional rotational symmetry in the $(y,z)$-plane.
Therefore, we can no longer decompose variables into scalar, vector and tensor components independently \footnote{In Ref.~\cite{Pereira:2007yy}, the SVT decomposition around Bianchi type I universe is performed in a time-dependent way.
Under this decomposition, the equations of motion for the scalar, vector and tensor variables become coupled.}.
Using the rotational symmetry in the $(y,z)$-plane, we can choose the wave vector of a Fourier mode as $\vec{k}=(k_x,k_y,0)$ without loss of generality. The background and this wave vector are invariant under the reflection
$(x,y,z)\rightarrow(x,y,-z)$. We therefore classify linear perturbations into even- and odd-parity sectors under this reflection, including the corresponding transformations of tensor indices. For a generic wave vector with $k_y\neq0$, these sectors coincide with the longitudinal (two-dimensional scalar) and transverse (two-dimensional vector) components in the $(y,z)$-plane, respectively. In this classification, the metric perturbations $\delta g_{\mu\nu}$ contain
\begin{align}
    &\text{even}: \delta g_{00}, \; \delta g_{0x}, \; \delta g_{xx}, \; 1 \; \text{of} \; \delta g_{0a}, \; 1 \; \text{of} \; \delta g_{xa}, \; 2  \; \text{of} \; \delta g_{ab}, \quad \text{odd}:  1 \; \text{of} \; \delta g_{0a}, \; 1 \; \text{of} \; \delta g_{xa}, \; 1  \; \text{of} \; \delta g_{ab}.
    \notag
    \end{align}
Here and in the following, the indices $(a, b)$ run over $(y, z)$.
For brevity, we write $k_{a}$ for the transverse component of the wave vector, so that $k^{2}_{a}=k^{2}_{y}$ for this choice, and we call a wave vector with $k_{a}\neq0$ generic.
Compared to the isotropic case, the two tensor polarizations of FLRW belong to different parity sectors: one is even and the other is odd. Likewise, a component belonging to the usual isotropic vector sector is classified as even in the present decomposition.

We now write the perturbed metric in a form that makes this decomposition manifest. Keeping only linear perturbations, we set
\begin{align}
    ds^2&=-(1+2A)
    \frac{dN^2}{H^2}+2e^{2\alpha-4\beta}B_{x}H^{-1}dNdx+2e^{2\alpha+2\beta}B_{a}H^{-1}dNdx^{a} \nonumber \\
    &+e^{2\alpha}\qty{e^{-4\beta}(1+2C-4D)dx^2+2e^{2\beta}E_{a}dxdx^{a} +e^{2\beta}\qty[(1+2C+2D)\delta_{ab}+2f_{ab}]dx^{a}dx^{b}}.
    \label{eq:metp}
    \end{align}
The metric and matter perturbations and their parity classification are summarized in Tab.~\ref{tab:vars}.
\begin{table}[t]
\centering
\begin{tabular}{cll}
\hline
Variable & Meaning & Parity \\
\hline
$A$ & lapse perturbation & even \\
$B_x$ & shift along $x$ & even \\
$B_a$ & shift in the $(y,z)$-plane & even ($\partial_a B$) + odd \\
$C$ & isotropic expansion perturbation & even \\
$D$ & anisotropic expansion perturbation & even \\
$E_a$ & off-diagonal $(x,a)$ component & even ($\partial_a E$) + odd \\
$f_{ab}$ & trace-free part in the $(y,z)$-plane & even ($f$) + odd \\
\hline
$\delta\phi$ & scalar-field perturbation & even \\
$\delta A_0$ & time component of $\delta A_\mu$ & even \\
$\delta A_x$ & $x$ component of $\delta A_\mu$ & even \\
$\delta A_a$ & $(y,z)$ components of $\delta A_\mu$ & even ($\partial_a\delta A$) + odd \\
\hline
\end{tabular}
\caption{Parity classification of the metric and matter perturbations for $\vec{k}=(k_x,k_y,0)$ under the reflection $z\rightarrow-z$.}
\label{tab:vars}
\end{table}
For the even-parity components of $B_a$, $E_a$, $f_{ab}$ and $\delta A_a$, we introduce the potentials $B$, $E$, $f$ and $\delta A$ via
\begin{align}
B^{(\mathrm{e})}_{a} = \partial_{a}B, \quad
E^{(\mathrm{e})}_{a} = \partial_{a}E, \quad
f^{(\mathrm{e})}_{ab}=\qty(\partial_{a}\partial_{b}-\frac{1}{2}\delta_{ab}\partial^2_{c})f, \quad
\delta A^{(\mathrm{e})}_{a} = \partial_{a}\delta A,
    \notag
\end{align}
where the superscript $(\mathrm{e})$ denotes even parity, while $(\mathrm{o})$ labels the odd-parity components constructed from transverse two-dimensional vectors.

In total, the even-parity sector contains 7 metric variables ($A, B_x, B, C, D, E, f$) and 4 matter variables ($\delta\phi, \delta A_0, \delta A_x, \delta A$), i.e., 11 even-parity perturbations. 
The odd-parity sector contains 3 metric variables (the transverse parts of $B_a, E_a, f_{ab}$) and 1 matter variable (the transverse part of $\delta A_a$), i.e., 4 odd-parity perturbations. 

In the following sections, we focus on the even-parity perturbations.
The odd-parity perturbations are discussed separately in App.~\ref{VP}.
For generic wave vectors with $k_a\neq0$, the six physical directions of the even-parity sector found in Sec.~\ref{directions} and the four physical integration constants of the odd-parity sector together account for ten initial data. 
They correspond to the five propagating degrees of freedom: the scalar field, the two physical polarizations of the gauge field, and two dynamical tensor degrees of freedom of the metric, which reduce to gravitational-wave polarizations in the isotropic limit.

\subsection{Even-parity perturbations and gauge fixing}
\label{SPG}

We now specialize to the even-parity sector and discuss its gauge structure.
We first specify the even-parity components of the metric and matter perturbations and then derive their transformation laws under infinitesimal spacetime and U(1) gauge transformations. 
We then fix the gauge conditions that will be used throughout most of this paper and identify the residual gauge freedoms that remain after this fixing.

The even-parity components are obtained from Eq.~\eqref{eq:metp} by setting $B_a=\partial_a B$, $E_a=\partial_a E$ and $f_{ab}=\qty(\partial_{a}\partial_{b}-\frac{1}{2}\delta_{ab}\partial^2_{c})f$, together with $\delta\phi$ and $\delta A_{\mu}=(\delta A_{0},\delta A_{x},\partial_{a}\delta A)$.
In what follows, we drop the superscript $(\mathrm{e})$ for notational simplicity.

We now turn to the gauge degrees of freedom.
Under an infinitesimal gauge transformation generated by $\delta x^{\mu}$, the metric perturbations transform as
\begin{align}
    \delta g_{\mu\nu}\rightarrow \delta g_{\mu\nu} - \bar{g}_{\alpha\nu}\partial_{\mu}(\delta x^{\alpha}) - \bar{g}_{\mu\alpha}\partial_{\nu}(\delta x^{\alpha})-\partial_{\alpha}(\bar{ g}_{\mu\nu})\delta x^{\alpha}.
    \end{align}
If we write the transformation by $\delta x^{\mu}=(n,L^{x},\delta^{ab}\partial_{b}L)$,
we obtain
\begin{align}
    &A\rightarrow A-n' + \frac{H'}{H}n, \quad B_{x}\rightarrow B_{x} - H L'^{x} +e^{-2\alpha+4\beta}\frac{\partial_{x}n}{H}, \quad B\rightarrow B - HL' + e^{-2\alpha-2\beta}\frac{n}{H}, 
    \notag
    \\
    &C\rightarrow C-\frac{1}{3}\qty(\partial_{x}L^{x}+\partial^2_{a}L)-n, \quad D\rightarrow D-\frac{1}{6}\qty(-2\partial_{x}L^{x}+\partial^2_{a}L)-\beta'n,  
    \notag
    \\ &E\rightarrow E-\partial_{x}L-e^{-6\beta}L^{x}, \quad 
    f\rightarrow f-L,
    \notag
    \end{align}
and
\begin{align}
    &\delta\phi \rightarrow \delta\phi-\phi'n, \quad \delta A_{0} \rightarrow \delta A_{0} - \bar{A}_{x}L'^x, \quad \delta A_{x} \rightarrow \delta A_{x} - \bar{A}_{x}\partial_{x}L^x - \bar{A}'_{x}n, \quad \delta A \rightarrow \delta A - \bar{A}_{x}L^x.
    \notag
\end{align}

Having established the transformation rules, we now specify the gauge conditions that will be primarily used in the rest of the paper.
The perturbed $e$-folding number is defined as
\begin{align}
    \mathcal{N}(N,{\bf x}) &:= \int^N  (1+A)\frac{\mathcal{K}}{3} \frac{dN_*}{H} \simeq 
    N + \frac{1}{3}\int^N \qty(3HC' -\partial_{x}B_x -\partial_{a}B_a ) \frac{dN_*}{H}
    \label{eq:mathN}
    \end{align}
with
\begin{align}
    \mathcal{K}:=3H\qty(1-A+C')-\partial_{x}B_x -\partial_{a}B_a 
    \label{eq:K}
\end{align}
being the trace of the extrinsic curvature on constant-time slices.
In the second equality in Eq.~\eqref{eq:mathN}, we have retained only terms up to linear order in the perturbations.

Our main choice is the ``uniform-$\mathcal{N}$ slicing'' condition, defined by the condition $\mathcal{N}({\bf x})=N$.
From Eq.~\eqref{eq:mathN}, this condition is equivalent to
\begin{align}
3HC' -\partial_{x}B_x -\partial_{a}B_a =0. \label{eq:Ng1}
    \end{align}
This fixes the time slicing while leaving the spatial threading free. 
To remove the remaining spatial gauge freedom, we additionally impose the vanishing shift condition $\delta g_{0i} = 0$.
With these choices, the gauge conditions for the geometry reduce to
\begin{align}
B_{x}=0, \quad B=0, \quad C'&=0. \label{eq:Ng2}
    \notag
    \end{align}
In the following, for convenience, we refer to this set of conditions as the ``$\mathcal{N}$-gauge" (uniform-$\mathcal{N}$ gauge).
This is the anisotropic analogue of the gauge $B=\dot H_L=0$ used in the isotropic case of Ref.~\cite{Sasaki:1998ug}.

For the vector field, there is a U(1) gauge degree of freedom.
We fix that U(1) gauge by imposing
\begin{align}
    \delta A_{0}=0.
\end{align}

We note that even after imposing the above conditions, residual gauge freedom remains for both spacetime and U(1).
The allowed residual transformations are
\begin{align}
    &x^{\mu}\rightarrow x^{\mu}+\delta x^{\mu}, \quad \delta x^{\mu} = \qty(n({\bf x}), L^{x}({\bf x}), \delta^{ab}\partial_{b}L({\bf x})), \label{eq:resgauge} \\
    &\delta A_{\mu}\rightarrow \delta A_{\mu}+\partial_{\mu}\theta({\bf x}). \label{eq:resu1gauge}
    \end{align}
These correspond to:
\begin{itemize}
    \item $n({\bf x})$: a shift of the background time $N$, which shifts $C$ by a constant, together with $D$, $\delta\phi$ and $\delta A_{x}$.
    \item $L^{x}({\bf x})$: spatial diffeomorphism along $x$, which shifts $E$ and $\delta A$, and through $\partial_{x}L^{x}$ also $C$, $D$ and $\delta A_{x}$.
    \item $L({\bf x})$: spatial diffeomorphism in the $(y, z)$-plane, which shifts $f$, and through $\partial^{2}_{a}L$ also $C$, $D$ and $E$.
    \item $\theta({\bf x})$: U(1) gauge transformation, which shifts $\delta A$ by $\theta$ and $\delta A_{x}$ by $\partial_{x}\theta$.
\end{itemize}
At leading order in the gradient expansion, these transformations preserve the gauge conditions.
For a spatially varying $n({\bf x})$, keeping $B_{x}=B=0$ requires $L'^{x}$ and $L'$ proportional to $\partial_{i}n$, which are of first order in spatial gradients and can be neglected at this order.
These transformations shift the constant parts of $E$, $f$ and $\delta A$, and shift the constant $C$ together with $D$, $\delta\phi$ and $\delta A_{x}$.
Which of these shifts are pure gauge, and how the residual freedom is fixed, is determined in Sec.~\ref{LWP} once the long-wavelength equations have been solved.
The roles of all perturbation variables are summarized in Tab.~\ref{tab:dof}.

\section{Long-wavelength perturbations and conserved quantities}
\label{LWP}

In this section, we analyze the linear perturbations on superhorizon scales using the gradient expansion method. 
In Sec.~\ref{PE}, we first introduce the gradient expansion and classify the perturbation variables. 
The two classes are those that can be absorbed into a redefinition of the background parameters (Sec.~\ref{Cor}), and those that are directly integrable in time (Sec.~\ref{SS}). 
In Sec.~\ref{directions}, we discuss which of the modes generated by the redefinition of the background parameters are pure gauge and which are physical.
In Sec.~\ref{W}, we combine the momentum constraints to obtain a quantity $W$ that is conserved on superhorizon scales and that will serve as a key tool for evaluating the curvature perturbation.
While the analysis in this section and the following is performed at linear order, the structural correspondence established here has a direct non-linear counterpart, which we discuss separately in App.~\ref{NL} within the ADM formalism.

\subsection{Gradient expansion and classification}
\label{PE}

For an isotropic background, the spatial gradient expansion assumes that spatial gradients of perturbed variables are small compared to their time derivatives, which are characterized by the Hubble expansion rate $H$.
In an anisotropic background, however, the expansion rate is direction dependent, $H_x = H(1-2\beta')$ along the $x$-direction and $H_a = H(1+\beta')$ in the $(y,z)$-plane, and so is the physical wavenumber.
We therefore introduce a single bookkeeping parameter $\epsilon$ defined by
\begin{align}
    \epsilon^2 := \frac{(k_x e^{-\alpha + 2\beta})^2 + (k_a e^{-\alpha - \beta})^2}{H^2},
    \end{align}
that is, we assume that the physical wavenumber along each direction is smaller than the Hubble rate.
During inflation where $H$ is nearly constant, $\epsilon$ decreases as long as both $e^{\alpha-2\beta}$ and $e^{\alpha+\beta}$ grow, which holds for $-1 < \beta' < 1/2$.
Thus $\epsilon \ll 1$ is preserved once the mode exits the horizon.

We expand the regular metric and matter perturbation components in powers of $\epsilon$, with the scalar potentials $E$, $f$, and $\delta A$ counted through their spatial derivatives.
The full set of linearized equations is given in App.~\ref{PEspacetime}, and their order-by-order structure in App.~\ref{DGE}.

At the leading order, spatial derivative terms are subleading and can be neglected, which significantly simplifies the system.
Among the perturbation equations in the superhorizon regime, two qualitatively different classes of variables emerge.
The classification is summarized in Tab.~\ref{tab:dof}.

\paragraph{Class I: Variables with background counterparts}

These variables obey equations that have a direct analogue in the background system. 
This class includes $(A, C, D, \delta\phi, \delta A_x)$.
They are treated in Sec.~\ref{Cor}.

\paragraph{Class II: Variables without background counterparts}

These variables have no counterpart in the background system within the axisymmetric ansatz~\eqref{eq:B1}. 
We note that the potentials $E$, $f$ and $\delta A$ may contain inverse powers of the wavenumber, e.g., $d_{E}\propto k_{x}/k^{2}_{a}$, whereas the physical components $\partial_{a}E$, $\partial_{a}\partial_{b}f$ and $\partial_{a}\delta A$ remain regular for a generic wave vector. 
The special direction $k_{a}=0$ is discussed in Sec.~\ref{W2}.
They satisfy equations that can be integrated directly in time. 
This class includes the even-parity variables $(E,f,\delta A)$, treated in Sec.~\ref{SS}, and the odd-parity perturbations, treated in App.~\ref{VP}.

\subsection{Class I: Variables with background counterparts} 
\label{Cor}

The Class I variables $(A,C,D,\delta\phi,\delta A_x)$ obey equations whose structure mirrors that of the background equations.
In this subsection, we show that these perturbations can be constructed directly by varying the background solutions.

\subsubsection{Correspondence with the background equations}
\label{sec:corr}

As we have seen in Sec.~\ref{BGparam}, the background solutions are characterized by the seven parameters $(N, \lambda^{p}, \bar{s})$, i.e., the six integration constants $\lambda^{J}$ and the dilatation.
Consider an infinitesimal variation $\Delta$ that shifts these degrees of freedom.

As an example, the $x$-component of the Maxwell equation for the perturbations reads
\begin{align}  
&\partial_{N}\qty[I^2e^{\alpha+4\beta}H\bar{A}'_{x}\qty(\frac{\delta A'_{x}}{\bar{A}'_{x}}+2\frac{\partial_{\phi}I}{I}\delta\phi-A+C+4D)]=0, \label{eq:PMaxxG}
     \end{align}
while the corresponding variation of the background Maxwell equation~\eqref{eq:usof} gives
\begin{align}
&\partial_{N}\qty[I^2e^{\alpha+4\beta}H\bar{A}'_{x}\qty(\frac{(\Delta \bar{A}_{x})'}{\bar{A}'_{x}}+2\frac{\partial_{\phi}I}{I}\Delta\phi+\frac{\Delta H}{H}+\Delta\alpha+4\Delta\beta)]=0. \label{eq:bpertmax}
\end{align}

Comparing these two equations, and similarly the remaining perturbation equations with the varied background equations listed in App.~\ref{app:LO}, we find that they are identical under the replacement
\begin{align}
    &A \leftrightarrow -\frac{\Delta H}{H},\quad
    C\leftrightarrow \Delta\alpha  ,\quad D\leftrightarrow \Delta\beta  ,\quad \delta\phi \leftrightarrow \Delta\phi ,\quad \delta A_{x} \leftrightarrow \Delta \bar{A}_{x}.
    \label{eq:Cor}
\end{align}
This correspondence implies that the set $(A, C, D, \delta\phi, \delta A_{x})$ satisfies exactly the same linearized equations as a variation of the background solution with respect to its parameters.
Thus, the separate universe picture of the isotropic case~\cite{Sasaki:1998ug} is generalized to our anisotropic setup. 
Each superhorizon-sized patch can be regarded as evolving as an independent Bianchi type I universe with slightly different parameters. 
Consequently, the perturbations of the dynamical fields $\delta\chi^{m}:=(D, \delta\phi, \delta A_{x})$ can be expressed as linear combinations of partial derivatives of the background fields $\chi^{m}=(\beta, \phi,\bar{A}_{x})$ with respect to the parameters $\lambda^{J}$ and the dilatation.

\begin{align}
    \delta\chi^{m} & = c^{J}\partial_{J}\chi^{m} + s\Delta_s \chi^{m} =
    c^{N}\partial_{N}\chi^{m} + c^{p}\partial_{\lambda^{p}}\chi^{m} + s\Delta_s \chi^{m},
    \label{eq:delchi}
\end{align}
with $c^J = (c^{N}, c^{p})$ and $s$ being constants and $\partial_{J}$ denotes the partial derivative with respect to $\lambda^{J}$.

We now integrate the Class I equations needed later. 
Their integration constants are immediately expressed in terms of the background parameters via the correspondence~\eqref{eq:delchi}.

The $x$-component of the Maxwell equation Eq.~\eqref{eq:PMaxxG} can be integrated to give
\begin{align}
    &\frac{\delta d_{A}}{d_{A}}=\frac{\delta A'_{x}}{\bar{A}'_{x}}+2\frac{\partial_{\phi}I}{I}\delta\phi-A+C+4D, \label{eq:defCx}
     \end{align}
with $\delta d_{A}$ being an integration constant.
The $(x,x)$ traceless component of the Einstein equations reads
\begin{align}    
&H\partial_{N}(HD')+3H^2D'-2\qty[
H\partial_{N}(H\beta')+3H^2\beta']A-H^2\beta'A'=\frac{1}{3}\delta\mathcal{V}. \label{eq:pertxxG}
    \end{align}
where the perturbed energy density of the vector field is defined as
\begin{align}
\delta\mathcal{V}
     &:=2\mathcal{V}\qty(\frac{\delta A'_{x}}{\bar{A}'_{x}}+\frac{\partial_{\phi}I}{I}\delta\phi-A-C+2D).  \label{eq:defdV} 
  \end{align}
Similarly, Equation~\eqref{eq:pertxxG} can be integrated as
\begin{align}
    &D'-\beta'A=\frac{1}{He^{3\alpha}}\qty[d_{D}+\frac{d_{A}}{3}\qty{\delta A_{x}+\qty(\frac{\delta d_{A}}{d_{A}}-3C)\bar{A}_{x}}],   \label{eq:dDsol}
    \end{align}
with $d_{D}$ being another integration constant.
This relation indicates that the $(x,x)$-component of the shear tensor is decomposed into the decaying term proportional to $e^{-3\alpha}$ and the terms sourced by the vector field.
Using Eq.~\eqref{eq:defCx}, the perturbed energy density of the vector field~\eqref{eq:defdV} can be written as
\begin{align}
\delta\mathcal{V}
    &=2\mathcal{V}\qty(\frac{\delta d_{A}}{d_{A}}-\frac{\partial_{\phi}I}{I}\delta\phi-2C-2D) =He^{-3\alpha}d_{A}\qty[\qty(\frac{\delta d_{A}}{d_{A}}-3C)\bar{A}'_{x}-A\bar{A}'_{x}+\delta A'_{x}]. 
  \end{align}
The second equality follows from the relation~\eqref{eq:nu2}.

Comparing the perturbed Maxwell solution~\eqref{eq:defCx} with the background solution~\eqref{eq:usof}, via the correspondence~\eqref{eq:Cor}, we find that $\delta d_{A}$ is precisely the variation of the background parameter $d_{A}$:
\begin{align}
\delta d_{A} = \Delta d_{A}=c^{J}\partial_{J}d_{A} + 2s d_{A}.
\label{eq:delcx}
\end{align}
Likewise, comparing the equation for the shear~\eqref{eq:dDsol} with the background solution~\eqref{eq:Acbeta} for $\beta'$, and using both the correspondence~\eqref{eq:Cor} and the result~\eqref{eq:delcx}, we obtain
\begin{align}
d_{D} &= c^{J}\qty(\partial_{J}d_{\beta}-3d_{\beta}\partial_{J}\alpha). \label{eq:dD}
\end{align}
The dilatation does not contribute to $d_{D}$, since the shifts of $d_{\beta}$ and $\alpha$ cancel.

\subsubsection{Gauge and physical directions in the parameter space}
\label{directions}

As we discussed above, the Class-I variables are generated by variations of the background solution with the seven parameters $(N, \lambda^{p}, \bar{s})$.
In this subsubsection we clarify which of these directions are pure gauge and which are physical.

The count of seven directions matches the number of integration constants of the Class-I system.
The three second-order equations~\eqref{eq:ApertphiG}, \eqref{eq:PMaxxG} and \eqref{eq:pertxxG} supply six integration constants for $(D, \delta\phi, \delta A_x)$, and the gauge condition $C'=0$ gives one more constant.
$A$ is fixed by the Hamiltonian constraint~\eqref{eq:pert00G}.
This parallels the $2n+1$ constants in the isotropic multi-field case, where one constant, associated with the lower limit of the time-integration of the curvature perturbation, appears in addition to the $2n$ constants of the scalar-field
perturbation. 

\paragraph{Time translation.}

Since $N$ is the time coordinate, $\alpha=N$ and hence $\partial_N\alpha=1$.
This means that the variation $\partial_N$ displaces the isotropic volume as well as the matter fields.
From Eq.~\eqref{eq:Cor}, we obtain
\begin{align}
(A, C, D, \delta\phi, \delta A_{x}) = c^N \qty(-\frac{H'}{H},1,\beta',\phi',\bar{A}'_{x}) \label{eq:dilN}
\end{align}
for the variation along $\partial_N$.
The residual gauge transformation~\eqref{eq:resgauge} with $n(\vec{x}) = - c^N$, applied to the null perturbation, reproduces \eqref{eq:dilN}. 
Hence $\partial_N$ is a pure gauge direction, and its amplitude $c^N$ must drop out of every gauge-invariant quantity.
It is removed below by fixing the residual time gauge.

\paragraph{Dilatations.}

The dilatation $\Delta_{s}$ of Sec.~\ref{BGparam} is generated by the residual spatial transformation $\delta x^i=-s x^i$.
Under the dilatation, we have
\begin{align}
(A, C, D, \delta\phi, \delta A_{x})_{s} := s(0,1,0,0,\bar{A}_{x}). \label{eq:dilV}
\end{align}

We note that $C$ is generated only by the time-translation and the dilatation modes:
\begin{align}
    C = c^N +s.
    \label{eq:Ccs}
\end{align}
This is compatible with the gauge condition $C' = 0$.

Of the remaining six directions $(c^{p},s)$, three ($\Delta_{s}$, $\Delta_{\sigma}$ and $\Delta_{\theta}$) are generated by global coordinate and U(1) transformations and are pure gauge in the homogeneous limit, while the other three, e.g., $(\phi'_{0}, d_{A}, d_{\beta})$, correspond to genuinely different homogeneous universes.
Together they span all six directions, since $\Delta_{\sigma}$ and $\Delta_{\theta}$ contain $\partial_{\beta_{0}}$ and $\partial_{\bar{A}_{x0}}$, respectively.
Whether the former three remain physical at finite wavenumber depends on the constraints and on the residual gauge freedom, both of which also act on the Class II variables of Sec.~\ref{SS}.
We therefore return to this question in Sec.~\ref{W2}, after the Class II sector and the constraints have been solved, where we show that all six directions are physical for a generic wave vector and treat the special direction $k_{a}=0$.

\paragraph{Gauge part and the dilatation mode.}
The pure gauge direction $\partial_{N}$ can be separated from the physical ones in a simple way.
Using Eq.~\eqref{eq:Ccs}, the general Class I solution~\eqref{eq:delchi} can be rearranged as
\begin{align}
\delta\chi^{m} = C\,\partial_{N}\chi^{m} - s\,\hat{\partial}_{N}\chi^{m} + c^{p}\partial_{\lambda^{p}}\chi^{m},
\qquad \hat{\partial}_{N}:=\partial_{N}-\Delta_{s}. \label{eq:delchidec}
\end{align}
The constant $C$ thus measures exactly the amplitude of the pure gauge part.
The direction $\hat{\partial}_{N}$ leaves $C$ unchanged, $\hat{\partial}_{N}\alpha=0$.
It is a time translation at fixed $e$-folding number, i.e., the time shift, which also shifts $\alpha$, followed by the dilatation that brings $\alpha$ back.
Since $\partial_{N}$ is pure gauge, $\hat{\partial}_{N}$ is gauge equivalent to $-\Delta_{s}$.
It thus represents the dilatation, and we refer to it as the dilatation mode.
From Eqs.~\eqref{eq:dilN} and \eqref{eq:dilV}, it generates
\begin{align}
(A, C, D, \delta\phi, \delta A_{x}) = \qty(-\frac{H'}{H},\ 0,\ \beta',\ \phi',\ \bar{A}'_{x}-\bar{A}_{x}). \label{eq:hatN}
\end{align}
The gauge part is removed by the residual time reparametrization $n=C$, which sets $C=0$; this is done in Sec.~\ref{SS}, after the spatial residual gauge has been fixed.
Then $c^{N}=-s$ by Eq.~\eqref{eq:Ccs}, and Eq.~\eqref{eq:delchidec} becomes
\begin{align}
\delta\chi^{m} = c^{N}\hat{\partial}_{N}\chi^{m} + c^{p}\partial_{\lambda^{p}}\chi^{m} =: c^{J}\hat{\partial}_{J}\chi^{m}, \qquad \hat{\partial}_{J}:=\qty(\hat{\partial}_{N}, \partial_{\lambda^{p}}),
\label{eq:delchiC0}
\end{align}
with the six physical parameters $(c^{N}, c^{p})$.
In the scalar-only case, $\Delta_{s}$ acts only on $\alpha$, so that $\hat{\partial}_{N}$ reduces to $\partial/\partial N$ acting on the scalar fields, and Eq.~\eqref{eq:delchiC0} coincides with the parametrization $\chi=c^{\alpha}\partial\phi/\partial\lambda^{\alpha}$ $(\lambda^{1}=N)$ of Ref.~\cite{Sasaki:1998ug}, with $C=0$ corresponding to $H_{L}=0$ there.
The mode $c^{N}\leftrightarrow c^{1}$ is then the adiabatic growing mode of Ref.~\cite{Sasaki:1998ug}, i.e., the solution $\chi_{(1)}=\phi_{N}$ with constant comoving curvature perturbation $\mathcal{R}_{c}=-c^{1}$, as opposed to the adiabatic decaying mode, whose amplitude is $c^{\alpha}W_{(\alpha)}$.
In the anisotropic case, the dilatation mode does not keep the comoving curvature perturbation constant (Sec.~\ref{sec:late}), and we use the name dilatation mode throughout.
As discussed in Sec.~\ref{sec:flat}, the parameters $(c^{N},c^{p})$ are fixed by superhorizon initial data, which may be obtained by matching to quantum mode functions evolved from horizon crossing.

\subsection{Class II: Variables without background counterparts}
\label{SS}

We now turn to the second class of variables, namely $(E,f,\delta A)$.
Their equations can be integrated directly in time.

From the $a$-components and the time-component of the Maxwell equations,
we obtain
\begin{align}  
&\partial_{N}\qty[I^2e^{\alpha+4\beta}H\bar{A}'_{x}\qty(e^{-6\beta}\frac{\delta A'}{\bar{A}'_{x}}-E)] = 0, \label{eq:PMaxaG} \\
&\partial_{x}\qty(\frac{\delta A'_{x}}{\bar{A}'_{x}}+2\frac{\partial_{\phi}I}{I}\delta\phi-A+C+4D)+ \partial^2_{a}\qty(e^{-6\beta}\frac{\delta A'}{\bar{A}'_{x}}-E) = 0, \label{eq:PMaxtG}
     \end{align}
and from the traceless components of the Einstein equations,
\begin{align}
    &H\partial_{N}(HE')+H^2\qty(3+6\beta')E'=-2H^2I^2e^{-2\alpha-2\beta}\bar{A}'_{x}\delta A', \label{eq:pertxaG} \\
&H\partial_{N}(Hf')+3H^2f'=0.  \label{eq:pertabG} 
    \end{align} 

First, we discuss the solution of the perturbed Maxwell equation.
Equation~\eqref{eq:PMaxaG} can be integrated to give
\begin{align}
    &\frac{d_{\perp}}{d_{A}}=e^{-6\beta}\frac{\delta A'}{\bar{A}'_{x}}-E,
    \label{eq:defCa}
     \end{align}
where $d_{\perp}$ is an integration constant.
From the time-component~\eqref{eq:PMaxtG}, we obtain the following relation:
\begin{align}
    &\partial_{x}\delta d_{A}+\qty(\partial^2_{y} + \partial^2_{z})d_{\perp} = 0,  \label{eq:Cgauge}
     \end{align}
which indicates that $\delta d_{A}$ and $d_{\perp} $ are not completely independent.

Equation~\eqref{eq:pertxaG} can be written as
\begin{align}    &\partial_{N}\qty(e^{3\alpha+6\beta}HE')=-2d_{A}\delta A'. 
    \end{align}

This can be integrated as
    \begin{align}
&e^{3\alpha+6\beta}HE' =  \widetilde{d}_{E}-2d_{A}\delta A. \label{eq:EpSol}
    \end{align}
    
Here,  $\widetilde{d}_E$ is an integration constant.
By substituting Eq.~\eqref{eq:defCa} into Eq.~\eqref{eq:pertxaG}, we obtain
\begin{align}
    &\partial_{N}(HE')+
    H\qty(3+6\beta')E'=-2e^{-3\alpha}\bar{A}'_{x}(d_{\perp} +d_{A}E).
    \end{align} 
This can be solved to give an expression that does not involve the matter field:
\begin{align}    
e^{6\beta}E&=  c_{E} -e^{6\beta}\frac{d_{\perp}}{d_{A}} + d_{E}I_{\beta} \label{eq:Esol2}   
    \end{align}
with $c_E$ and $d_E$ being integration constants.
Here, we have defined
\begin{align}
 I_{\beta}(N) := \int^N_{N_{0}} e^{-3\alpha+
 6\beta}\frac{dN_*}{H}.
    \end{align}
In Eq.~\eqref{eq:Esol2}, the term $-d_{\perp}/d_{A}$ is time-independent.
It represents the transverse component of the electric flux, i.e., the direction of the electric field lines relative to the background $x$-axis.
The value of $d_{\perp}/d_{A}$ alone depends on the choice of spatial coordinates.
Under the residual transformation $L$ of Eq.~\eqref{eq:resgauge}, $E$ is shifted by $-\partial_{x}L$, so that $d_{\perp}/d_{A}$ is shifted by $\partial_{x}L$, while $\delta d_{A}$ is shifted by $-d_{A}\partial^{2}_{a}L$ through the shifts of $C$ and $D$.
These shifts leave Gauss's law~\eqref{eq:Cgauge} invariant.
The combination $d_{\perp}/d_{A}+E=e^{-6\beta}\delta A'/\bar{A}'_{x}$ is invariant under this residual $L$ transformation, although it is not invariant under every residual spatial transformation.

This solution has a simple geometric interpretation for $d_{E}=0$.
Consider a Bianchi type I patch whose anisotropy axis and gauge field are both rotated by a small angle $\theta_{a}=\partial_{a}\Theta$.
The rotation generates the off-diagonal metric component $\delta g_{xa}=e^{2\alpha}(e^{-4\beta}-e^{2\beta})\theta_{a}$, which gives $E=(e^{-6\beta}-1)\Theta$.
Comparing this with Eq.~\eqref{eq:Esol2}, the coefficient of $e^{-6\beta}$ and the constant term must agree separately, which requires $c_{E}=\Theta$ and $d_{\perp}/d_{A}=\Theta$.
The rotation angle thus determines both $c_{E}$ and $d_{\perp}/d_{A}$, and the rotated gauge field reproduces the solution for $\delta A$ obtained below.
Since $c_{E}$ can be shifted by the residual transformation $L^{x}$, a solution with $c_{E}\neq d_{\perp}/d_{A}$ corresponds to the rotated patch combined with a residual coordinate transformation.
In particular, after we set $c_{E}=0$ below, only the time-independent term $E=-d_{\perp}/d_{A}$ remains for $d_{E}=0$.

The remaining terms in Eq.~\eqref{eq:Esol2}, proportional to $c_{E}$ and $d_{E}$, encode the effects of the anisotropic expansion.
While the part proportional to $c_{E}$ can be removed by a residual gauge transformation as shown below, the $d_{E}$ part represents time evolution of $E$ induced by the anisotropic expansion.
Thus, the solution~\eqref{eq:Esol2} decomposes the perturbation $E$ into a persistent, conserved transverse flux and a dynamical shear-induced mode that captures the memory of the anisotropic expansion.

Using Eqs.~\eqref{eq:defCa} and \eqref{eq:Esol2}, the perturbation of the vector field is expressed as
\begin{align}
    &\delta A' = \bar{A}'_{x} \qty(c_{E} + d_{E}I_{\beta})
         \end{align}
 and this is integrated to yield
 \begin{align}        
    & \therefore \delta A =  c_{\delta A} + c_{E}\bar{A}_{x} + d_{E}I_{A\beta}(N).
        \label{eq:delAsol} 
     \end{align}
Here, $c_{\delta A}$ is a constant of integration, and we have introduced
\begin{align}
    &I_{A\beta}(N) := \int^N_{N_{0}} \bar{A}'_{x}  (N_{*})I_{\beta}(N_{*})dN_{*}.
     \end{align}

In Eq.~\eqref{eq:pertabG}, there is no source and we can straightforwardly integrate to obtain
\begin{align}
    f
    &=c_{f}+d_{f}I_{\alpha}(N),
    \label{eq:Fsol} 
    \end{align}
where $c_{f}$ and  $d_{f}$ are integration constants.
    
We also note that the integrability of Eqs.~\eqref{eq:pertxxG}, \eqref{eq:pertxaG} and \eqref{eq:pertabG} stems from the shear symmetry of the background
whose associated Noether charge $Q^{i}_{\, j}$ was identified in Ref.~\cite{Tanaka:2021dww}.
For the even-parity perturbation, the components of the Noether charge are identified as
\begin{align}
    &Q^{x}_{\, x} \leftrightarrow \text{}\delta d_{\beta},\quad
    Q^{x}_{\, a} \leftrightarrow \tilde{d}_E,\quad \left.Q^{a}_{\, b}\right|_{\textrm{traceless}}  \leftrightarrow d_f. \label{eq:NQ}
\end{align}
where
\begin{align}
&\delta d_{\beta}:= d_{D}+3d_{\beta}C. \label{eq:Ddef}
\end{align}
This combination is the variation of the background charge $d_{\beta}$ including the change of the local volume, $\delta\sqrt{g}=3C\sqrt{g}$.
In the gauge $C=0$ fixed at the end of this subsection, it reduces to $d_{D}$.
Using the background-perturbation correspondence~\eqref{eq:dD} together with $C = c^J \partial_J \alpha + s\Delta_s \alpha$ from~\eqref{eq:Cor}, the quantity $\delta d_{\beta}$ is immediately interpreted as the variation of the background parameter $d_{\beta}$:
\begin{align}
\delta d_{\beta} &= c^{J}\partial_{J}d_{\beta} + 3sd_{\beta} . \label{eq:dD2}
\end{align}
Furthermore, the momentum constraints, discussed in Sec.~\ref{W}, give the relation between the spatial gradient of some components of $Q^{i}_{\, j}$ and the matter fields.

The integration constants $c_{E}$, $c_{f}$ and $c_{\delta A}$  correspond to residual gauge degrees of freedom listed in Tab.~\ref{tab:dof}.
To eliminate these unphysical degrees of freedom, we perform a residual gauge transformation~\eqref{eq:resgauge} with
\begin{align}
 &L({\bf x}) = c_{f}, \quad L^{x}({\bf x}) = c_{E}. \label{eq:LLx}
    \end{align}

Under this transformation, the solutions simplify to
\begin{align}    
E&= -\qty(\frac{d_{\perp}}{d_{A}} + \partial_xc_{f}) + e^{-6\beta} d_{E}I_{\beta}(N), \quad f=d_{f}I_{\alpha}(N).
    \end{align}
Therefore, under the redefinition $d_{\perp}/d_{A} + \partial_xc_{f} \rightarrow d_{\perp}/d_{A}$, $c_{E}$ and $c_{f}$ can be removed.
Under the transformation~\eqref{eq:resgauge} with Eq.~\eqref{eq:LLx} and the residual U(1) gauge transformation~\eqref{eq:resu1gauge} with $\theta =  -c_{\delta A}$, the perturbed vector field can be written as
\begin{align}
    &\delta A = d_{E}I_{A\beta}(N) ,
        \label{eq:delAsol2} 
     \end{align}
where we have used Eq.~\eqref{eq:delAsol}.
The transformation also shifts the Class I variables and, in turn, the parameters $(c^{p},s)$; this is made explicit below.

Although the Class II variables have no counterpart within the axisymmetric ansatz~\eqref{eq:B1}, their solutions also admit a background interpretation.
At leading order in the gradient expansion, Eqs.~\eqref{eq:PMaxaG}--\eqref{eq:pertabG} coincide with the linearized equations for spatially homogeneous off-diagonal components of the metric and the gauge field around the background, i.e., with those of a general, non-axisymmetric Bianchi type I universe.
Accordingly, the integration constants fall into two groups.
The constants $c_{E}$, $c_{f}$ and $c_{\delta A}$, as well as the part of $d_{\perp}$ that can be shifted by $L$, are generated by the residual coordinate and U(1) transformations of the background.
In particular, the rotation discussed below Eq.~\eqref{eq:Esol2} is the combination of the residual transformations with $L^{x}=-\Theta$ and $\partial_{x}L=\Theta$.
The constants $d_{E}$ and $d_{f}$, on the other hand, are the variations of the off-diagonal shear charges of a general Bianchi type I background, i.e., of the off-diagonal components of the Noether charge $Q^{i}{}_{j}$ of Ref.~\cite{Tanaka:2021dww}.
For $f$, this is the variation of $d_{\gamma}$ in Eq.~\eqref{eq:Abeta2eq}.
All leading-order solutions, including those of the Class II sector, are therefore generated by variations of general Bianchi type I backgrounds and by coordinate transformations, in accordance with the anisotropic separate-universe picture of Ref.~\cite{Tanaka:2021dww}.

\paragraph{Complete fixing of the residual gauge.}
It remains to fix the residual time reparametrization $n$.
After the spatial transformation~\eqref{eq:LLx}, which shifts $C$ as above, we choose $n$ such that
\begin{align}
C=0. \label{eq:C0}
\end{align}
This removes the pure gauge term $C\,\partial_{N}\chi^{m}$ in Eq.~\eqref{eq:delchidec}.
The residual gauge $(n, L^{x}, L, \theta)$ is thus completely fixed by the conditions $c_{E}=c_{f}=c_{\delta A}=0$ and $C=0$, at leading order in the gradient expansion, as discussed in Sec.~\ref{SPG}. 
Since $n$ is determined only after $L^{x}$ and $L$, no ordering problem arises.
Similarly, the U(1) transformation $\theta=-c_{\delta A}$ shifts $\delta A_{x}$ by $-\partial_{x}c_{\delta A}$, i.e., along $\Delta_{\theta}$, so that the $\Delta_{\theta}$ component of $c^{p}$ also refers to this gauge.
In this gauge, $c^{N}=-s$, and the Class I solution takes the form~\eqref{eq:delchiC0} with the six physical parameters $(c^{N}, c^{p})$.
A time-independent transformation with $a:=\partial_{x}L^{x}$ and $b:=\partial^{2}_{a}L$ shifts $(C, D, \delta A_{x})$ by $(-(a+b)/3,\ (2a-b)/6,\ -a\bar{A}_{x})$, i.e., along $\Delta_{s}$ and $\Delta_{\sigma}$ with $s=-(a+b)/3$ and $\sigma=(2a-b)/6$.
The spatial transformation~\eqref{eq:LLx}, together with the compensating time reparametrization $n=-(a+b)/3$, therefore shifts $c^{N}$ by $(a+b)/3$ and the $\Delta_{\sigma}$ component of $c^{p}$ by $(2a-b)/6$.
The values of $(c^{N}, c^{p})$ refer to this gauge.
From Sec.~\ref{W} on, we set $C=0$ in all equations, whereas the equations of Secs.~\ref{PE}--\ref{SS} and App.~\ref{PEspacetime} are written for general $C$.
In the rest of this paper, we denote the transformed parameters by the same symbols.

The integration constant $\widetilde{d}_E$ in Eq.~\eqref{eq:EpSol} is related to $d_E$ as
\begin{align}
\widetilde{d}_E & =d_{E} e^{6\beta(N_0)} , \label{eq:dEtilde}
\end{align}
where $N_{0}$ is the lower limit of the integrals introduced in Eq.~\eqref{eq:Ia}.
The derivation is given in App.~\ref{app:dE}.

\begin{table}[t]
\centering
\begin{tabular}{c|c|c|c}
\hline
Class & Perturbative variable & Background counterpart~\eqref{eq:Cor} & Role \\
\hline
\multirow{5}{*}{I}
 & $\delta\phi$  & $\Delta\phi$     & Dynamical \\
 & $\delta A_x$  & $\Delta \bar{A}_{x}$     & Dynamical \\
 & $D$           & $\Delta\beta$    & Dynamical \\
 & $A$           & $-\Delta H/H$    & Constraint (Hamiltonian) \\
 & $C$           & $\Delta\alpha$   & $C=c^N+s=0$ (residual gauge) \\
\hline
\multirow{3}{*}{II}
 & $E$           & ---              & Fixed by constraints + gauge (space) \\
 & $f$           & ---              & Fixed by constraints + gauge (space) \\
 & $\delta A$    & ---              & Fixed by constraints + gauge (U(1)) \\
\hline
\end{tabular}
\caption{Summary of even-parity perturbations in the $\mathcal{N}$-gauge.
The variables with background counterparts (Class I) are treated through the separate-universe correspondence in Sec.~\ref{Cor}, while the remaining variables (Class II) are solved by direct integration in Sec.~\ref{SS}.
The dynamical variables obey evolution equations such as \eqref{eq:ApertphiG}, \eqref{eq:PMaxxG}, and \eqref{eq:pertxxG}. 
The constraint variable is determined by the Hamiltonian constraint~\eqref{eq:pert00G}.
The Class-II variables have no counterpart within the axisymmetric background~\eqref{eq:B1}.
Their constant parts ($c_E$, $c_f$, $c_{\delta A}$) are residual gauge modes. The remaining integration constants $d_E$ and $d_f$ are associated with the shear Noether charges of Ref.~\cite{Tanaka:2021dww}, whereas $d_{\perp}$ describes the transverse electric flux. These constants are fixed in terms of the Class~I sector by Gauss's law~\eqref{eq:Cgauge} and the momentum constraints derived below.}
  \label{tab:dof}
\end{table}

\subsection{Momentum constraints and conserved quantities}
\label{W}

In Secs.~\ref{Cor} and \ref{SS}, we obtained solutions for the perturbations in the Class I and Class II variables, respectively.
However, these two sets of solutions are not independent, but they are related by the momentum constraints.
In this subsection, we combine these constraints to derive a conserved quantity $W$ on superhorizon scales.

As shown in App.~\ref{DGE}, the momentum constraints and the Gauss constraint are not independent of the Hamiltonian constraint, the evolution equations and the matter-field equations within the gradient expansion. Once the latter are satisfied, the former are preserved in time.
Therefore, at leading order, it suffices to impose them at the initial time, where they reduce to algebraic relations among the integration constants of the leading-order solutions.
The momentum constraints read
\begin{align}
&-2H(1+\beta')\partial_{x}A+6H\beta'\partial_{x}D+2H\partial_{x}D'
-\frac{H}{2}\qty(\partial^2_{a}E'+6\beta'\partial^2_{a}E)=J_{x}, \label{eq:pert0xG} \\
    &H\partial_{a}\qty[-(2-\beta')A+6\beta'D-D']-H\partial_{b}f'_{ab}-\frac{H}{2}e^{6\beta}\partial_{x}\partial_{a}E'=J_{a}. \label{eq:pert0aG} 
\end{align}

\subsubsection{The conserved quantity $W$}
\label{W1}

We introduce the following combination of the Class I variables:
\begin{align}
    &W:=He^{3\alpha}\qty(A-3\beta'D-\frac{1}{2}\phi'\delta \phi)+\frac{\delta d_{A}}{6}\bar{A}_{x}-\frac{1}{3}d_{A}\delta A_{x}.
    \label{eq:Wdef}
\end{align}
A Wronskian-type representation of $W$ is given in App.~\ref{app:Wronskian}.

As shown in App.~\ref{app:Wcons}, $W'=0$ follows from the trace part of the spatial Einstein equations~\eqref{eq:perttrG}, the integrated Maxwell equation~\eqref{eq:defCx}, and the background equations~\eqref{eq:ABGphi}, \eqref{eq:ABG00}, \eqref{eq:ABGtr} and \eqref{eq:Abeta1eq}.
This indicates that $W$ remains frozen on superhorizon scales, for arbitrary $U(\phi)$ and $I(\phi)$ within this model.

The quantity $W$ is a combination of metric perturbations $(A,D)$, the scalar field perturbation $\delta\phi$, and gauge field perturbations $(\delta d_{A}, \delta A_x)$.
That is to say, $W$ is composed only of the perturbative variables belonging to Class I.
Then, using the relation~\eqref{eq:Cor}, we can express $W$ as
\begin{align}
    &W=He^{3\alpha}\qty(-\frac{\Delta H}{H}-3\beta' \Delta\beta - \frac{1}{2}\phi'\Delta \phi)+\frac{ \Delta d_{A}}{6}\bar{A}_{x}-\frac{1}{3}d_{A}\Delta \bar{A}_{x}.
\end{align}
In particular, using the background equations~\eqref{eq:ABG00}, \eqref{eq:ABGtr} and \eqref{eq:nu2}, we can show that $W$ vanishes for the time-translation mode.
Furthermore, from Eqs.~\eqref{eq:dD2} and \eqref{eq:delcx}, we see that $\delta d_{\beta}$ and $\delta d_{A}$ also vanish for this mode.
Since $W$ also vanishes for the dilatation~\eqref{eq:dilV}, it vanishes for the dilatation mode $\hat{\partial}_{N}$ of Eq.~\eqref{eq:hatN}.
In contrast, $\delta d_{\beta}$ and $\delta d_{A}$ do not vanish for $\hat{\partial}_{N}$, as shown in Sec.~\ref{W2}.

Let us consider the behavior of $W$ in the isotropic limit, which we define by switching off the background anisotropic sources, $\qty(d_{\beta} \rightarrow0, d_{A}\rightarrow 0)$, while keeping the perturbations of the background parameters fixed in relative form.
In particular, in this limit, we fix the ratio $\delta d_{A}/d_{A}$, so that $\delta d_{A} \rightarrow 0$ and $d_{A} \rightarrow 0$.
Since $He^{3\alpha}\beta' = d_{\beta}+d_{A}\bar{A}_{x}/3$ from Eq.~\eqref{eq:Acbeta}, this is equivalent to $\beta'\rightarrow0$ and $d_{A}\rightarrow0$.
In the isotropic limit, 
it reduces to the familiar conserved quantity in the isotropic case~\cite{Sasaki:1998ug}:
\begin{align}
    &W_{\textrm{iso}}=He^{3\alpha}\qty(A-\frac{1}{2}\phi'\delta \phi).
\end{align}

We stress that this isotropic limit is taken in formulas derived for $d_{A}\neq0$, with $\delta d_{A}/d_{A}$ held fixed, and differs from the scalar-only case of Sec.~\ref{BG}, in which $d_{A}$, $\delta d_{A}$ and $d_{\perp}$ are not defined and the flux terms of Sec.~\ref{R and G} are absent.
In the scalar-only case, the shear-charge variation $\delta d_{\beta}$ decouples from the scalar-field curvature evolution, while the even-parity sector still contains a tensor degree of freedom.
The quantity $W_{\textrm{iso}}$ is precisely the conserved Wronskian for multi-scalar model in an FLRW background. 
Thus, our anisotropic separate-universe framework provides a genuine generalization that smoothly connects to the well-established isotropic results in the absence of anisotropic sources.

In terms of these quantities, the momentum constraints take a simple form.
Substituting the solutions \eqref{eq:dDsol}, \eqref{eq:EpSol}, and \eqref{eq:Fsol} into the momentum constraints~\eqref{eq:pert0xG} and \eqref{eq:pert0aG}, and using the relation~\eqref{eq:Cgauge}, we obtain

\begin{align}
&\partial^2_{a}d_{E}=\partial_{x}\qty(-4W +4\delta d_{\beta}-6d_{\beta}\frac{\delta d_{A}}{d_{A}}), \label{eq:dtheta} \\
&d^{f}_{ab,b}=
    \partial_{a}\qty(-2W-\delta d_{\beta}-\frac{1}{2}\partial_{x}\widetilde{d}_{E}) = \partial_{a}\qty(-2W-\delta d_{\beta}) -\frac{e^{6\beta(N_0)}}{2}\partial_{a}\partial_{x}d_{E}. \label{eq:df}
\end{align}
Here, $d^{f}_{ab}:=\qty(\delta^{ac}\partial_{c}\partial_{b}-\frac{1}{2}\delta^{a}_{\, b}\partial^2_{c})d_f$.

\subsubsection{Independence of $(W, \delta d_{\beta}, \delta d_{A})$}
\label{W2}

For any Class I solution~\eqref{eq:delchiC0}, $W$ is a linear function of the parameters,
\begin{align}
W = c^{N}W_{\hat{N}} + c^{p}W_{p} = c^{p}W_{p}, \label{eq:WJ}
\end{align}
where $W_{\hat{N}}$ and $W_{p}$ denote the values of $W$ on the modes generated by $\hat{\partial}_{N}$ and $\partial_{\lambda^{p}}$, respectively, and we have used $W_{\hat{N}}=0$ shown in Sec.~\ref{W1}.
This is the counterpart of $W=c^{\alpha}W_{(\alpha)}$ with $W_{(1)}=0$ in Ref.~\cite{Sasaki:1998ug}.
From Eqs.~\eqref{eq:dD} and \eqref{eq:dD2}, we see that $\delta d_{A}$ and $\delta d_{\beta}$ are nothing but the variations of the background integration constants $d_{A}$ and $d_{\beta}$, respectively.
Moreover, since these are constants of the background motion, they are transverse to the time-translation direction $\partial_{N}$ (but not to $\hat{\partial}_{N}$).
Concretely, we can express them as
\begin{align}
   \delta d_{A}= c^{d_{A}} - 2c^{N} d_{A}, \quad \delta d_{\beta} = c^{d_{\beta}} - 3c^{N} d_{\beta}. \label{eq:dhatN}
\end{align}
We can show that $W$ is independent of $\delta d_{A}$ and $\delta d_{\beta}$ as follows.
Let us consider the mode generated by $\partial_{\phi'_{0}}$, i.e., the variation of $\phi'$ on the slice $\phi = \phi_0$ with the other parameters being fixed.
As $W$ is a constant, it is sufficient to show only at one time-slice. 
This mode lies in the kernel of both $\delta d_{A}$ and $\delta d_{\beta}$.
Under this variation, we have $W = - e^{3\alpha}\Delta H$.

Since the right-hand side of Eq.~\eqref{eq:Hrecon} is unchanged under this variation, varying Eq.~\eqref{eq:Hrecon} gives $\Delta H = H\phi'\Delta\phi'/(6-\phi'^2)$. 
Hence, for $\Delta\phi'=1$, we have
\begin{align}
   W & = - \frac{e^{3\alpha} H \phi'}{6 - \phi'^2} \neq 0
\end{align}
for this mode.
Hence the map $(c^{N}, c^p)$ $\rightarrow (W, \delta d_{\beta}, \delta d_{A})$ is rank three, since $c^{d_{A}}$ and $c^{d_{\beta}}$ enter $\delta d_{A}$ and $\delta d_{\beta}$ independently.
This holds even for $d_{\beta}=d_{A}=0$, although the explicit expressions containing $1/d_{A}$ and $1/\bar{A}'_{x}$, such as Eqs.~\eqref{eq:Cgauge} and \eqref{eq:dtheta}, apply only for a nonvanishing background electric field. 
What changes in the scalar-only case is not the rank, but the way these data enter the curvature: the shear-charge variation cancels from the scalar-field curvature evolution, and the electric-flux fluctuation decouples at linear order. 
This is distinct from the isotropic limit of Sec.~\ref{W1}, as discussed in Sec.~\ref{sec:late}.

Since $W_{\phi'_{0}}\neq0$, by reparametrizing the background parameters as
\begin{align}
   \tilde{\lambda}^{q} := \lambda^{q} \quad (\lambda^{q}\neq\phi'_{0}), \qquad \tilde{\phi}'_{0} := \phi'_{0} + \sum_{\lambda^{q}\neq\phi'_{0}}\frac{W_{q}}{W_{\phi'_{0}}}\lambda^{q},
\end{align}
we can choose the basis such that $W_{q} = 0$ for all $\lambda^{q}\neq\tilde{\phi}'_{0}$.
In this basis, the direction $\tilde{\phi}'_{0}$ corresponds to the adiabatic decaying mode in Ref.~\cite{Sasaki:1998ug}. 

In summary, the physical integration constants of the Class II sector, $(d_{E}, d_{f}, d_{\perp})$, are expressed through Eqs.~\eqref{eq:Cgauge}, \eqref{eq:dtheta} and \eqref{eq:df} in terms of the three independent quantities $W$, $\delta d_{\beta}$, and $\delta d_{A}$, that can be expressed as the variation of the background quantities.
The Class I sector itself is characterized by the six parameters $(c^{N}, c^{p})$. Besides the three conserved combinations $(W, \delta d_{\beta}, \delta d_{A})$, the remaining directions enter the curvature directly through $D$, as shown in Sec.~\ref{RN}.
Together, these data characterize the superhorizon evolution of the curvature perturbation and the tensor metric perturbation discussed in Sec.~\ref{R and G}.

\paragraph{Physical independence.}
We now show that $(W,\delta d_{\beta},\delta d_A)$ provide independent conserved quantities in the residual-gauge convention fixed in Sec.~\ref{SS}.

First, the constraints do not relate them to each other.
The Gauss constraint~\eqref{eq:Cgauge} and the momentum constraints~\eqref{eq:dtheta} and \eqref{eq:df} determine the Class II constants $(d_{\perp}, d_{E}, d_{f})$ in terms of the gradients of $(W, \delta d_{\beta}, \delta d_{A})$, but impose no relation among the latter for a generic wave vector.
The only exception is $k_{a}=0$, where they require $\delta d_{A}=0$ and $W=\delta d_{\beta}$, as discussed below.

Second, none of the three quantities can be removed by a gauge transformation, since the residual gauge freedom has been completely used to set $c_{E}=c_{f}=c_{\delta A}=0$ and $C=0$ in Sec.~\ref{SS}.
Before this fixing, the time-independent residual transformations shift the Class I variables along $\Delta_{s}$, $\Delta_{\sigma}$ and $\Delta_{\theta}$, and hence the values of these quantities, but only together with the constants $c_{E}$, $c_{f}$ and $c_{\delta A}$; they leave the Noether charges $d_{E}$ and $d_{f}$ invariant and cannot generate the time dependence through $I_{\alpha}(N)$ and $I_{\beta}(N)$.

Third, the three quantities give distinguishable contributions to the curvature.
As we will see in Sec.~\ref{RN}, they enter the spatial curvature evaluated there through the combinations $2W+\delta d_{\beta}$, $-4W+4\delta d_{\beta}-6d_{\beta}\delta d_{A}/d_{A}$ and $\delta d_{A}/d_{A}$, which multiply $e^{-2\beta}k^{2}_{a}I_{\alpha}$, $e^{-2\beta}k^{2}_{x}(I_{\beta}-e^{6\beta(N_{0})}I_{\alpha}/2)$ and $e^{4\beta}k^{2}_{x}$, respectively.
The linear map from $(W, \delta d_{\beta}, \delta d_{A})$ to these combinations has the determinant $12/d_{A}\neq0$, and the three coefficient functions depend differently on time and on the direction of $\vec{k}$.
The three channels can therefore, in principle, be distinguished by the time and angular dependence of the curvature, although the shear $D$ contributes to the same curvature.

We refer to $W$, $\delta d_{\beta}$ and $\delta d_{A}$ as the scalar-field, shear and electric-flux channels, respectively.
This naming depends on the origin of $\bar{A}_{x}$. Under $\bar{A}_{x}\rightarrow\bar{A}_{x}+c$, we have $\delta d_{\beta}\rightarrow\delta d_{\beta}-c\,\delta d_{A}/3$ and $W\rightarrow W+c\,\delta d_{A}/6$.
It therefore refers to the convention $\bar{A}_{x0}=0$ of Sec.~\ref{BGparam}, whereas the three combinations above are independent of this choice.

\paragraph{Physical directions and the special direction $k_{a}=0$.}
\label{ka0}
We can now settle the question left open in Sec.~\ref{directions}.
In the homogeneous limit, $\Delta_s$ and $\Delta_\sigma$ are mere rescalings of the spatial coordinates, and $\Delta_{\theta}$ of Eq.~\eqref{eq:Dtheta} is a global U(1) transformation.
For long-wavelength profiles $s(\vec x)$, $\sigma(\vec x)$ and $\bar{\theta}(\vec x)$, however, they are physical; for $\bar{\theta}(\vec x)$, e.g., $\delta A_{x}=\bar{\theta}$ and $\delta A=0$ give the long-wavelength magnetic field $F_{xa}=-\partial_{a}\bar{\theta}$.
At finite wavenumber with $k_{a}\neq0$, the residual gauge freedom has been completely used in Sec.~\ref{SS}, and the constraints~\eqref{eq:Cgauge}, \eqref{eq:dtheta} and \eqref{eq:df} fix the Class II constants without restricting the Class I sector.
Hence, among the seven directions of Sec.~\ref{directions}, only $\partial_{N}$ is pure gauge, and all six directions $(c^{N},c^{p})$ are physical for a generic wave vector.
Being pure gauge in the homogeneous limit does not by itself make $\Delta_{s}$, $\Delta_{\sigma}$ and $\Delta_{\theta}$ Weinberg adiabatic modes~\cite{Weinberg:2003sw} in the sense of Ref.~\cite{Tanaka:2021dww}, i.e., long-wavelength limits of physical solutions obtained from a residual gauge transformation: the constraints in general require Class II constants that do not vanish as $k\rightarrow0$, so that the extension to finite wavenumber is not generated by the gauge transformation alone.

For $\vec{k}$ parallel to the background vector field ($k_{a}=0$), the counting differs.
The reflection plane used to define the parity sectors is not unique, and the potentials $E$, $f$ and $\delta A$ are not defined, since their physical components $\partial_{a}E$, $\partial_{a}\partial_{b}f$ and $\partial_{a}\delta A$ vanish identically.
The residual transformations $L^{x}(x)$ and $\theta(x)$ are then no longer used to remove $c_{E}$ and $c_{\delta A}$, but act on the Class I sector: $L^{x}(x)$, combined with the compensating time reparametrization that keeps $C=0$, shifts the parameters along a combination of $\hat{\partial}_{N}$ and $\Delta_{\sigma}$, and $\theta(x)$ shifts them along $\Delta_{\theta}$.
The constraints, in turn, impose $\delta d_{A}=0$ and $W=\delta d_{\beta}$ on the Class I sector, and both residual directions preserve these two relations.
Hence, of the four parameters that satisfy the constraints, only two are physical.
This agrees with the field content for $\vec{k}\parallel\bar{\vec{A}}$, where the residual SO(2) symmetry about $\vec{k}$ leaves a single propagating helicity-zero degree of freedom, the scalar field.
In the following, the main results are stated for a generic wave vector, and we refer to this paragraph for the special direction.

\paragraph{Dilatation mode and the conserved charges.}
Equation~\eqref{eq:dhatN} shows the main physical difference from the isotropic case.
In the isotropic case, the corresponding adiabatic growing mode does not contribute to any decaying channel, $W_{(1)}=0$.
Here $W_{\hat{N}}=0$ as well, but the dilatation mode carries the variations $\delta d_{\beta}=-3c^{N}d_{\beta}$ and $\delta d_{A}=-2c^{N}d_{A}$ of the background charges.
Since $\hat{\partial}_{N}$ shifts the clock of the fields at fixed volume, it rescales the shear and the electric flux per comoving volume.
Through the momentum constraints, these variations feed into the Class II sector and, as we will see in Sec.~\ref{RN}, into the memory term of the curvature, with coefficients proportional to the background charges $d_{\beta}$ and $d_{A}$.

\section{Spatial curvature and tensor metric perturbations}
\label{R and G}

For single-field attractor inflation, the comoving curvature perturbation is conserved on superhorizon scales, whereas in multi-field models it can evolve through non-adiabatic contributions.
For isotropic multi-field models, Ref.~\cite{Sasaki:1998ug} expressed this evolution in terms of homogeneous background variations and a conserved Wronskian.
In the anisotropic model, the gauge field affects the intrinsic curvature of uniform-$\mathcal N$ hypersurfaces and couples the even-parity metric and matter perturbations.
This motivates two questions: (i) How can the intrinsic curvature on uniform-$\mathcal N$ hypersurfaces be evaluated explicitly from background solutions? (ii) How is the resulting curvature related to a metric perturbation that reduces to the plus-polarized tensor mode in the isotropic limit?

In Sec.~\ref{RN}, we evaluate the spatial curvature in the $\mathcal{N}$-gauge.
In Sec.~\ref{Rc}, we introduce the comoving slicings, and in Sec.~\ref{sec:flat}, we express the initial-condition parameters in terms of the flat-slicing variables.
The curvature perturbation at the end of the anisotropic stage is discussed in Sec.~\ref{sec:late}, and its relation to a tensor metric perturbation in Sec.~\ref{GW}.

\subsection{Spatial curvature in $\mathcal{N}$-gauge}
\label{RN}

We first evaluate the intrinsic curvature of constant-time slices in the $\mathcal{N}$-gauge.
The intrinsic curvature on a constant-time slice is given by
\begin{align}
    \frac{e^{2\alpha}}{2}R_{\mathcal{N}}&=e^{-2\beta}\qty[\partial_{a}\partial_{b}f_{ab}+\partial^2_{a}(-2C+D)]+e^{4\beta}\partial_{x}\partial_{a}E_{a}-2e^{4\beta}\partial^2_{x}(C+D) \nonumber \\
    &=-2\Lop  C-2\LopD{}D+e^{4\beta}\partial_{x}\partial^2_{a}E+\frac{e^{-2\beta}}{2}\partial^2_{a}\partial^2_{b}f, \label{eq:R}
    \end{align}
where we have introduced
\begin{align}    \Lop:=e^{4\beta}\partial^2_{x}+e^{-2\beta}\partial^2_{a}, \quad \LopD:=e^{4\beta}\partial^2_{x}-\frac{e^{-2\beta}}{2}\partial^2_{a}.
\end{align}  
Here and in the following, $R$ denotes the Ricci scalar of the spatial hypersurface, which should be distinguished from the curvature-perturbation amplitudes $\mathcal{R}$ introduced in Sec.~\ref{sec:flat}, which correspond to the curvature perturbation used in Ref.~\cite{Sasaki:1998ug}.

For a Fourier mode with wave vector $\vec{k}=(k_x,k_a)$, these operators become
\begin{align}
\Lopk&:=-\qty(e^{4\beta}k^2_{x}+e^{-2\beta}k^2_{a}), \quad 
\LopDk:=-\qty(e^{4\beta}k^2_{x}-\frac{e^{-2\beta}}{2}k^2_{a}).
    \end{align}
By substituting the superhorizon solutions for $E$ and $f$, Eqs.~\eqref{eq:Esol2} and~\eqref{eq:Fsol}, we obtain
\begin{align}
    \frac{e^{2\alpha}}{2}R_{\mathcal{N}}
    &=-2\Lopk{}C-2\LopDk{}D+e^{4\beta}ik_{x}k^2_{a}\frac{d_{\perp}}{d_{A}}
    -e^{-2\beta}ik_{x}k^2_{a}d_{E}I_{\beta}
+\frac{e^{-2\beta}}{2}k^2_{a}k^2_{b}d_{f}I_{\alpha}.
    \label{eq:Rkefs}
    \end{align}
Here, $c_E$ and $c_f$ have been removed by the residual gauge transformation~\eqref{eq:LLx}.
Furthermore, by using Eqs.~\eqref{eq:Cgauge}, ~\eqref{eq:dtheta} and \eqref{eq:df}, we obtain an expression for the curvature in terms of $W$, $\delta d_{\beta}$, $\delta d_{A}$ and the Class I variables:
\begin{align}
    \frac{e^{2\alpha}}{2}R_{\mathcal{N}}
    &=-2\LopDk{}D
    \nonumber \\
   &-e^{4\beta}k^2_{x}\frac{\delta d_{A}}{d_{A}}-e^{-2\beta}
   k^2_{x}\qty(-4W +4\delta d_{\beta}-6d_{\beta}\frac{\delta d_{A}}{d_{A}})\qty(- \frac{e^{6\beta(N_0)}}{2}I_{\alpha} +I_{\beta})+e^{-2\beta}k^2_{a}\qty(2W+\delta d_{\beta})I_{\alpha} .
    \label{eq:RW}
    \end{align}
Here we have set $C=0$ by Eq.~\eqref{eq:C0}. For general $C$, the term $-2\Lopk C$ is added to the right-hand side.
In this expression, the terms proportional to $I_{\alpha}(N)$ and $I_{\beta}(N)$ represent the integrated history of the expansion. 
In contrast, the term proportional to $\delta d_{A}$, which we call the flux term, appears without time integration, multiplied by the anisotropic scale factor $e^{4\beta}$. 
This reflects the fact that a gradient of the flux $\delta d_{A}$ requires a transverse flux $d_{\perp}$, i.e., a change of the direction of the electric field lines, through Eq.~\eqref{eq:Cgauge}, which directly induces a spatial anisotropy in the curvature on superhorizon scales without cumulative time evolution.

Equation~\eqref{eq:RW} expresses the intrinsic curvature in terms of the generalized Wronskian $W$, the shear- and electric-flux-charge variations $(\delta d_{\beta},\delta d_A)$, and the metric anisotropy $D$. The Class II integration constants have been eliminated using Gauss's law and the momentum constraints. The term $\LopDk D$ represents a direct contribution of the anisotropic spatial-metric perturbation, which has no counterpart in the scalar-field sector analyzed in Ref.~\cite{Sasaki:1998ug}.

Using Eqs.~\eqref{eq:delchiC0}, \eqref{eq:WJ} and \eqref{eq:dhatN}, the curvature is decomposed as
\begin{align}
R_{\mathcal{N}} = R_{\mathcal{N}}^{[\hat{N}]} + \sum_{p}R_{\mathcal{N}}^{[p]},
\end{align}
where $R_{\mathcal{N}}^{[\hat{N}]}$ is generated by the dilatation mode $\hat{\partial}_{N}$ and $R_{\mathcal{N}}^{[p]}$ by $\partial_{\lambda^{p}}$.
For the dilatation mode, Eqs.~\eqref{eq:hatN} and \eqref{eq:dhatN} give
\begin{align}
\frac{e^{2\alpha}}{2}R_{\mathcal{N}}^{[\hat{N}]} = \qty[2e^{4\beta}k^{2}_{x} - 2\beta'\LopDk - 3e^{-2\beta}k^{2}_{a}d_{\beta}I_{\alpha}]c^{N}. \label{eq:RhatN}
\end{align}
The first term in Eq.~\eqref{eq:RhatN} is the flux term, which arises from the transverse flux required by Gauss's law for $\delta d_{A}=-2c^{N}d_{A}$.
In the scalar-only case, this term is absent, and with $\beta'=d_{\beta}=0$ the whole expression vanishes, in agreement with the isotropic solution $\chi_{(1)}=\phi_{N}$ with $\mathcal{R}=0$.
In the limit $d_{A}\rightarrow0$ with $\delta d_{A}/d_{A}$ fixed, adopted in Sec.~\ref{W1}, the flux term instead remains finite, since $\delta d_{A}/d_{A}=-2c^{N}$ does not depend on the strength of the background field.
This non-uniformity of the limit reflects the fact that the direction of the electric field lines, $d_{\perp}/d_{A}$, which enters $E$ through Eq.~\eqref{eq:defCa}, remains finite as the field strength vanishes, so that fixing $\delta d_{A}/d_{A}$ fixes a geometric deformation of the slices rather than a field amplitude.
In the scalar-only case, $d_{\perp}/d_{A}$ is not defined and the term is absent.
In the anisotropic case, the dilatation mode contributes to the curvature through three effects. These are the shear $D=\beta'c^{N}$, the transverse flux induced by $\delta d_{A}=-2c^{N}d_{A}$ through Gauss's law, which gives the first term, and the memory term proportional to $d_{\beta}I_{\alpha}$ induced by $\delta d_{\beta}=-3c^{N}d_{\beta}$.
The terms $R_{\mathcal{N}}^{[p]}$ contain contributions proportional to the integrals $I_{\alpha}(N)$ and $I_{\beta}(N)$, which represent a memory of the anisotropic expansion that can persist even after the vector field decays.
Since the residual time gauge has been fixed, $R_{\mathcal{N}}$ is no longer ambiguous under time reparametrizations. Without the condition~\eqref{eq:C0}, it would contain the additional gauge-dependent contribution $-2(\Lopk+\beta'\LopDk)c^{N}$ of the time-translation mode, as shown in Sec.~\ref{Rc}.
The split between $R_{\mathcal{N}}^{[\hat{N}]}$ and the $\Delta_{\sigma}$ component of $\sum_{p}R_{\mathcal{N}}^{[p]}$ refers to the spatial gauge fixed by Eq.~\eqref{eq:LLx} in Sec.~\ref{SS}, while the total $R_{\mathcal{N}}$ is invariant under spatial diffeomorphisms because the background slices are flat.

\subsection{Comoving slicings}
\label{Rc}

In isotropic cosmology, a commonly used choice is the comoving slicing, defined by the vanishing of the scalar momentum density. 
In the present anisotropic system, however, the momentum current has distinct components parallel and perpendicular to the preferred direction of the background vector field. 
Therefore, there is no unique condition that can be referred to as “the” comoving slicing during the anisotropic phase. 
Instead, we consider two natural directional choices, defined by $J_x = 0$ and $J_a = 0$, respectively.
These are distinct directional prescriptions and do not generally impose the vanishing of every component of $J_i$ on a single hypersurface.

Under the gauge transformation of the time, $N\rightarrow N +dN$, the spatial curvature and the perturbation of the matter fields vary as
\begin{align}
    \frac{e^{2\alpha}}{2}R_{\mathcal{N}} \rightarrow  \frac{e^{2\alpha}}{2}R_{\mathcal{N}} + 2\qty(\Lop  +\beta'\LopD)dN, \label{eq:RdN} \\
    \delta\phi \rightarrow  \delta\phi -\phi' dN, \quad \delta A_{x} \rightarrow \delta A_{x} - \bar{A}'_{x}dN.
    \end{align}

In order to impose the comoving condition along the $x$-direction $\qty(J_{x}=0)$, we should take the gauge transformation as
\begin{align}
    dN_{C,(x)} : = \frac{\delta \phi}{\phi'}.
    \end{align}
On the other hand, the comoving condition along the $a$-direction $\qty(J_{a}=0)$ yields
\begin{align}
    dN_{C,(a)} : = \frac{\phi'\delta \phi + e^{-2\alpha+4\beta}I^2 \bar{A}'_{x}\qty(\delta A_{x} - \partial_{x}\delta A)}{\phi'^2 + e^{-2\alpha+4\beta}I^2(\bar{A}'_{x})^2}.
    \end{align}
Using these conditions, the spatial curvature on the comoving slices along the $x$-direction and $a$-direction are given as
\begin{align}
    &R_{C,(x)} := R_{\mathcal{N}} + \frac{4}{e^{2\alpha}}\qty(\Lop  +\beta'\LopD)dN_{C,(x)},
    \quad    
R_{C,(a)} := R_{\mathcal{N}} + \frac{4}{e^{2\alpha}}\qty(\Lop  +\beta'\LopD)dN_{C,(a)}, \label{eq:Rc_def}
    \end{align}
respectively.

Their difference reflects the preferred direction of the background vector field, not a gauge ambiguity. 
For $k_{x}=0$, $J_{x}$ vanishes identically, so that only $R_{C,(a)}$ is defined. Likewise, only $R_{C,(x)}$ is defined for $k_{a}=0$.
In the scalar-only case, both conditions reduce to the same $dN_{C,(a)} = dN_{C,(x)} =  \delta \phi/\phi'$, recovering the standard comoving slicing.
These two directional curvature perturbations are evaluated at the end of the anisotropic stage in Sec.~\ref{sec:late}.

\subsection{Flat slicing and initial conditions}
\label{sec:flat}

Another useful choice is the flat slicing, defined here by the vanishing of the intrinsic three-dimensional Ricci scalar at linear order; the full intrinsic Ricci tensor need not vanish.
From Eq.~\eqref{eq:RdN}, the required time shift is
\begin{align}
dN_{F} := -\frac{e^{2\alpha}}{4}\qty(\Lop  +\beta'\LopD)^{-1}R_{\mathcal{N}}, \label{eq:dNF}
    \end{align}
For nonzero Fourier modes, the inverse operator is defined by division by its eigenvalue; the homogeneous mode is not fixed by the spatial-curvature condition.
Then, the matter fields on the flat slicing are
\begin{align}
    &\delta\phi_{F} := \delta\phi - \phi'dN_{F}, \quad \delta A_{x,F} := \delta A_{x} - \bar{A}'_{x}dN_{F}.
    \end{align}
Similarly, we define the anisotropy on the flat slicing by
\begin{align}
    &D_{F} := D - \beta'dN_{F},
    \end{align}
which is slice-invariant.

More generally, we denote by $\delta\hat{\chi}_{F}^{m} := \qty(\delta\phi_{F}, \delta A_{x,F}, D_{F},\delta\phi'_{F}, \delta A'_{x,F}, D'_{F})$, and also write as $\hat{\chi}^{m} := \qty(\phi, \bar{A}_{x}, \beta, \phi', \bar{A}'_{x}, \beta')$, where the last three components of $\delta\hat{\chi}_{F}^{m}$ denote the $N$ derivatives of the first three, e.g., $\delta\phi'_{F}:=\partial_{N}(\delta\phi_{F})$. 
For nonzero Fourier modes, the flat-slicing condition fixes the time slicing, while the residual spatial and U(1) gauges are fixed as in Sec.~\ref{SS}. 
The variables $\delta\hat{\chi}_{F}^{m}$ are then invariant under linear time-slicing transformations and provide the physical initial data in this gauge convention.
In the gauge~\eqref{eq:C0}, it is convenient to express the time shift $dN_{F}$ as a curvature perturbation of the $\mathcal{N}$-slices,
\begin{align}
\mathcal{R}_{\mathcal{N}} := dN_{F}, \qquad\text{i.e.,}\qquad R_{\mathcal{N}} = -\frac{4}{e^{2\alpha}}\qty(\Lop+\beta'\LopD)\mathcal{R}_{\mathcal{N}}. \label{eq:calRN}
\end{align}
In the isotropic case, this reduces to the relation ${}^{(3)}R=-(4/a^{2})\nabla^{2}\mathcal{R}$, and $\mathcal{R}_{\mathcal{N}}$ to the curvature perturbation $\mathcal{R}$ in the gauge $B=\dot H_{L}=0$ with $H_{L}=0$.
From Eq.~\eqref{eq:RW}, we have
\begin{align}
\mathcal{R}_{\mathcal{N}} = \qty(\Lopk+\beta'\LopDk)^{-1}\qty[\LopDk D - \frac{1}{2}\mathcal{T}], \label{eq:calRNsol}
\end{align}
where $\mathcal{T}$ denotes the terms in Eq.~\eqref{eq:RW} other than that proportional to $D$, i.e., the flux term and the memory terms.
In the scalar-only case and for $c^{\beta_{0}}=0$, Eq.~\eqref{eq:calRNsol} reduces to $\mathcal{R}_{\mathcal{N}}=W I_{\alpha}=c^{p}W_{p}\int dN/(He^{3\alpha})$, since the contribution of $\delta d_{\beta}$ cancels against that of the decaying part of $D$.
This is precisely Eq.~(2.27) of Ref.~\cite{Sasaki:1998ug}, $\mathcal{R}=c^{\alpha}W_{(\alpha)}\int dN/(a^{3}H)$. 
The case $c^{\beta_{0}}\neq0$ is discussed in Sec.~\ref{sec:late}.

In the long-wavelength limit, Eq.~\eqref{eq:delchiC0} gives $\delta\hat{\chi}^{m}=c^{J}\hat{\partial}_{J}\hat{\chi}^{m}$ in the $\mathcal{N}$-gauge.
Since $\delta\phi_{F}=\delta\phi-\phi'\mathcal{R}_{\mathcal{N}}$ and similarly for the other fields, while the velocities also receive the derivative of the time shift, $\delta\phi'=\delta\phi'_{F}+\phi''\mathcal{R}_{\mathcal{N}}+\phi'\mathcal{R}'_{\mathcal{N}}$, we have
\begin{align}
\delta\hat{\chi}_{F}^{m} + \mathcal{R}_{\mathcal{N}}\,\partial_{N}\hat{\chi}^{m} + \mathcal{R}'_{\mathcal{N}}\,v^{m} = c^{J}\hat{\partial}_{J}\hat{\chi}^{m}, \qquad v^{m}:=\qty(0,0,0,\phi',\bar{A}'_{x},\beta') \quad (m=1,...,6). \label{eq:chiF}
\end{align}
In the isotropic case, $\mathcal{R}_{\mathcal{N}}\partial_{N}\phi = W\psi$ with $\psi$ defined in Ref.~\cite{Sasaki:1998ug}, and Eq.~\eqref{eq:chiF} reduces to their Eq.~(2.34), $\chi_{F}=c^{\alpha}(\chi_{(\alpha)}-W_{(\alpha)}\psi)$, together with its $N$ derivative.

The time shift $\mathcal{R}_{\mathcal{N}}$ is not an independent input but is itself a linear combination of the parameters, $\mathcal{R}_{\mathcal{N}}=r_{J}c^{J}$.
The coefficients $r_{J}$ are the values of $\mathcal{R}_{\mathcal{N}}$ on the modes $\hat{\partial}_{J}$. 
They follow from Eqs.~\eqref{eq:RW} and \eqref{eq:calRNsol} with the values of $(W, \delta d_{\beta}, \delta d_{A}, D)$ on each mode, and are therefore determined by the background solutions alone.
Since the integrals $I_{\alpha}$ and $I_{\beta}$ vanish at their lower limit $N_{0}$, the memory terms vanish there, so that $r_{J}(N_{0})$ is given by the shear and flux terms, and $r'_{J}(N_{0})$ follows from the $N$ derivative of Eq.~\eqref{eq:calRNsol}.
Equation~\eqref{eq:chiF} then reads $\delta\hat{\chi}_{F}^{m}=\mathcal{J}_{\rm flat}{}^{m}{}_{J}\,c^{J}$, and we obtain
\begin{align}
    &c^{J} = \qty[\qty(\mathcal{J}_{\rm flat}^{-1})^{J}{}_{m}\,\delta\hat{\chi}_{F}^{m}]_{N_0}, \qquad
    \mathcal{J}_{\rm flat}{}^{m}{}_{J} := \hat{\partial}_{J}\hat{\chi}^{m} - r_{J}\,\partial_{N}\hat{\chi}^{m} - r'_{J}\,v^{m},
    \label{eq:ci}
\end{align}
where $\hat{\partial}_{J}\hat{\chi}^{m}$ is the Jacobian matrix
\begin{align}
    &\mathcal{J} := \qty(\hat{\partial}_{N}\hat{\chi}^{m}, \partial_{\phi'_{0}}\hat{\chi}^{m}, \partial_{\beta_{0}}\hat{\chi}^{m}, \partial_{\bar{A}_{x0}}\hat{\chi}^{m}, \partial_{d_{A}}\hat{\chi}^{m}, \partial_{d_{\beta}}\hat{\chi}^{m}).
\end{align}
Here, we have assumed that $\mathcal{J}_{\rm flat}$ is invertible.
We identify the reference time $N_{0}$ of Eq.~\eqref{eq:Ia} with the initial time of the superhorizon evolution.
It should be chosen shortly after the mode satisfies $\epsilon\ll1$, i.e., a few $e$-folds after horizon exit, where the leading-order gradient expansion applies.
At horizon crossing, $\epsilon\simeq1$, and gradient corrections must be taken into account when matching to the quantum fluctuations.
Such matching must also respect the spatial and U(1) gauge convention used to define the superhorizon initial data.

\paragraph{Classical and quantum inputs.}
The procedure separates into two parts.
The matrix $\mathcal{J}_{\rm flat}$, the conserved quantities $(W,\delta d_{\beta},\delta d_A)$, the memory integrals $I_{\alpha}$ and $I_{\beta}$, and hence the map from $(c^{N},c^{p})$ to the curvature perturbation and the metric amplitude $G$
at $N_e$, are determined by the classical background dynamics.
In particular, $\mathcal{J}$ consists of derivatives of background solutions with respect to their parameters and, together with $r_{J}$, can be obtained, e.g., by finite differences of numerically solved backgrounds, as in the $\delta N$ formalism.
The only input from the subhorizon quantum theory is the set of initial values $\delta\hat{\chi}^{m}_{F}$ at $N_{0}$, i.e., $\delta\phi_{F}$, $\delta A_{x,F}$ and $D_{F}$ together with their time derivatives.
These are read off a few $e$-folds after horizon exit from the mode functions of the scalar field, of the gauge field with the coupling $I(\phi)$, and of the metric perturbation $D_{F}$, which reduces to one tensor polarization in the isotropic limit.
The integration constants of the Class II sector require no further input, since they are fixed by the constraints~\eqref{eq:Cgauge}, \eqref{eq:dtheta} and \eqref{eq:df} up to the residual gauge.
Given these mode functions, the computation of the curvature perturbation at $N_{e}$ thus reduces to the inversion of a $6\times6$ matrix built from background solutions.
Since $r_{J}$ depends on the direction of $\vec{k}$ through $e^{6\beta}k^{2}_{x}/k^{2}_{a}$, this matrix is to be evaluated for each direction of the wave vector.
For $k_{a}=0$, the constraints impose the two relations $\delta d_{A}=0$ and $W=\delta d_{\beta}$ on the Class I sector, so that the initial data must satisfy them, and only two physical parameters remain after the residual gauge is removed (Sec.~\ref{W2}).
Equation~\eqref{eq:ci} is the anisotropic generalization of the standard formula in isotropic multi-scalar model~\cite{Sasaki:1998ug}, Eqs.~(2.51)--(2.52) there, with $\mathcal{J}_{\rm flat}$ corresponding to their $\hat{\chi}^{A}_{(\alpha)}-W_{(\alpha)}\hat{\psi}^{A}$.
It shows explicitly that the six physical parameters $(c^{N}, c^{p})$ are completely determined by the initial values of $\delta\hat{\chi}_{F}^{m}$, i.e., the flat-slicing variables.
In particular, as we saw in the previous section, the conserved quantities $W, \delta d_{\beta}$ and $\delta d_{A}$ are linear combinations of these $(c^{N}, c^{p})$.
Thus, they are determined solely by the six physical initial-condition parameters $(c^{N}, c^{p})$.

\subsection{Curvature perturbation at the end of the anisotropic stage}
\label{sec:late}

Our superhorizon solutions apply as long as the background is described by the model~\eqref{eq:S} and $\epsilon\ll1$.
We therefore evaluate the curvature perturbation at a time $N_{e}$ at the end of the anisotropic stage, e.g., at the end of inflation, where the parameters $(c^{N}, c^{p})$ are related to the initial data by Eq.~\eqref{eq:ci}.
At this time the comoving slicing is still direction dependent, and we use the two directional curvature perturbations of Sec.~\ref{Rc}.
Defining their amplitudes by $R_{C,(i)}=-(4/e^{2\alpha})(\Lop+\beta'\LopD)\mathcal{R}_{C,(i)}$ $(i=x,a)$ as in Eq.~\eqref{eq:calRN}, Eqs.~\eqref{eq:Rc_def} and \eqref{eq:calRN} give
\begin{align}
\mathcal{R}_{C,(x)} &= \mathcal{R}_{\mathcal{N}} - \frac{\delta\phi}{\phi'}, \label{eq:RCx}\\
\mathcal{R}_{C,(a)} &= \mathcal{R}_{\mathcal{N}} - \frac{\phi'\delta\phi+e^{-2\alpha+4\beta}I^{2}\bar{A}'_{x}\qty(\delta A_{x}-\partial_{x}\delta A)}{\phi'^{2}+e^{-2\alpha+4\beta}I^{2}(\bar{A}'_{x})^{2}}, \label{eq:RCa}
\end{align}
to be evaluated at $N=N_{e}$.
Equation~\eqref{eq:RCx} has the same form as Eqs.~(2.44)--(2.45) of Ref.~\cite{Sasaki:1998ug}, $\mathcal{R}_{c}=\mathcal{R}+\Delta N$ with $\Delta N=-\phi_{N}\cdot\chi/\phi_{N}^{2}$.
Substituting $\delta\phi=c^{N}\phi'+c^{p}\partial_{\lambda^{p}}\phi$ from Eq.~\eqref{eq:delchiC0}, we obtain
\begin{align}
\mathcal{R}_{C,(x)} = -c^{N} - c^{p}\frac{\partial_{\lambda^{p}}\phi}{\phi'} + \mathcal{R}_{\mathcal{N}}, \label{eq:RCxdec}
\end{align}
which corresponds term by term to their Eq.~(2.46), $\mathcal{R}_{c}=-c^{\alpha}\phi_{N}\cdot\chi_{(\alpha)}/\phi_{N}^{2}+c^{\alpha}W_{(\alpha)}\int dN/(a^{3}H)$.
The first term is the dilatation mode, which corresponds to their adiabatic growing mode with $c^{N}\leftrightarrow c^{1}$.
The second term is the contribution of the other background parameters.
The third term is the intrinsic-curvature contribution on uniform-$\mathcal N$ hypersurfaces. 
It contains both flux and memory terms in an anisotropic background and reduces to $c^{p}W_{p}\int dN/(He^{3\alpha})$ in the scalar-only case, as follows from Eq.~\eqref{eq:calRNsol}.

To make the comparison explicit, we separate from $\mathcal{R}_{\mathcal{N}}$ the part that has the isotropic form, $\mathcal{R}_{\mathcal{N}} = WI_{\alpha} + \Delta\mathcal{R}_{\rm ani}$ with $W=c^{p}W_{p}$.
Equation~\eqref{eq:RCxdec} then becomes
\begin{align}
\mathcal{R}_{C,(x)} = -c^{N} - c^{p}\qty(\frac{\partial_{\lambda^{p}}\phi}{\phi'} - W_{p}I_{\alpha}) + \Delta\mathcal{R}_{\rm ani}. \label{eq:RCxani}
\end{align}
The first two terms have the form of Eq.~(2.46) of Ref.~\cite{Sasaki:1998ug} with $c^{N}\leftrightarrow c^{1}$ and $\lambda^{p}\leftrightarrow\lambda^{\alpha\geq2}$, while $\Delta\mathcal{R}_{\rm ani}$ collects the explicit additional terms. 
The coefficients of the first two terms can nevertheless depend on the anisotropic background evolution.
The split between the first two terms and $\Delta\mathcal{R}_{\rm ani}$ depends on the origin of $\bar{A}_{x}$ through $W$, while their sum does not. The dependence disappears in the isotropic limit, where $\delta d_{A}\rightarrow0$.
From Eqs.~\eqref{eq:RW} and \eqref{eq:calRNsol}, we find
\begin{align}
\Delta\mathcal{R}_{\rm ani} &= \qty(\Lopk+\beta'\LopDk)^{-1}\Bigg\{\LopDk\qty(D-\delta d_{\beta}I_{\alpha}-\beta'WI_{\alpha}) \nonumber\\
&\qquad + e^{-2\beta}k^{2}_{x}\qty[\mathcal{I}_{\Delta}\qty(W-\delta d_{\beta}) + \frac{1}{2}\qty(e^{6\beta}+3d_{\beta}\qty(\mathcal{I}_{\Delta}-e^{6\beta}I_{\alpha}))\frac{\delta d_{A}}{d_{A}}]\Bigg\}, \label{eq:DRani}\\
\mathcal{I}_{\Delta} &:= \qty(e^{6\beta}+e^{6\beta(N_{0})})I_{\alpha} - 2I_{\beta}. \nonumber
\end{align}
Here $D-\delta d_{\beta}I_{\alpha}$ is the shear with its memory part due to the shear charge removed, so that the cancellation between $\delta d_{\beta}$ and the decaying part of $D$ found in the isotropic case is built in.
The function $\mathcal{I}_{\Delta}$ vanishes identically when $\beta=0$ and hence measures the difference between the memory integrals along $k_{x}$ and $k_{a}$.
The last term describes the transverse flux and its coupling to the background shear.

In what follows, we express the direction dependence through
\begin{align}
\mu^{2}:=\frac{e^{4\beta}k^{2}_{x}}{e^{4\beta}k^{2}_{x}+e^{-2\beta}k^{2}_{a}}, \label{eq:mu}
\end{align}
the squared cosine of the angle between the physical wave vector and the $x$-axis.
In terms of $\mu$, we have $\Lopk=-k^{2}_{\rm p}$ and $\LopDk=-k^{2}_{\rm p}P_{2}(\mu)$, where $k^{2}_{\rm p}:=e^{4\beta}k^{2}_{x}+e^{-2\beta}k^{2}_{a}$ is the squared physical wavenumber up to the factor $e^{-2\alpha}$, and $P_{2}(\mu)=(3\mu^{2}-1)/2$.
For fixed comoving $\vec{k}$, the physical direction evolves according to $(\mu^2)'=6\beta'\mu^2(1-\mu^2)$.

In the last term of Eq.~\eqref{eq:DRani}, the part proportional to $e^{6\beta}$ is the flux term, while the part proportional to $d_{\beta}$ arises from the momentum constraint~\eqref{eq:dtheta}.
In the scalar-only case with the background shear switched off ($d_{\beta}=\beta'=0$), the flux term is absent. 
With $\mathcal{I}_{\Delta}=0$ and $D=c^{\beta_0}+\delta d_{\beta}I_{\alpha}$, we have
\begin{align}
\Delta\mathcal{R}_{\rm ani} \rightarrow \Lopk^{-1}\LopDk\,c^{\beta_{0}} = P_{2}(\mu)\,c^{\beta_{0}}. \label{eq:DRiso}
\end{align}
In this limit, Eq.~\eqref{eq:Dsigma} gives $\Delta_{\sigma}=\partial_{\beta_{0}}$, so that this term is the contribution of the anisotropic rescaling.
In this case, $\Delta_{\sigma}$ shifts the traceless part of the spatial metric by a constant and contains a long-wavelength tensor-mode component, although this spatial-metric perturbation is not itself transverse-traceless for a general direction of $\vec{k}$. 
It belongs to the even-parity sector of Sec.~\ref{Dec}.
It has no counterpart in the analysis of Ref.~\cite{Sasaki:1998ug}. For $c^{\beta_{0}}=0$, Eq.~\eqref{eq:RCxani} reduces exactly to their Eq.~(2.46).
The absence of the flux term refers to the scalar-only case; in the isotropic limit of Sec.~\ref{W1}, the last term of Eq.~\eqref{eq:DRani} survives.
For $k_{a}=0$, Eq.~\eqref{eq:DRani} should not be split into the contributions of the individual modes $(c^{N}, c^{p})$, since the amplitudes $c^{N}$ and $c^{d_{A}}$ are then not independent (Sec.~\ref{W2}).

The physical difference from the isotropic case appears in $\Delta\mathcal{R}_{\rm ani}$. 
While the dilatation mode contributes only the constant $-c^{1}$ in the isotropic case, here it also contributes to $\mathcal{R}_{\mathcal{N}}$ through Eq.~\eqref{eq:RhatN}, as discussed in Sec.~\ref{W}.
In terms of $\mu$, Eqs.~\eqref{eq:RhatN} and \eqref{eq:calRNsol} give
\begin{align}
\mathcal{R}^{[\hat{N}]}_{\mathcal{N}}=c^{N}\,\frac{2\mu^{2}+\beta'(3\mu^{2}-1)-3(1-\mu^{2})\,d_{\beta}I_{\alpha}}{2+\beta'(3\mu^{2}-1)}, \label{eq:RhatNmu}
\end{align}
where the three terms in the numerator arise from the flux term, the shear $D=\beta'c^{N}$ and the memory term, respectively.
For nearly constant $H$ and $\beta'$, as in the attractor solution of Ref.~\cite{Watanabe:2009ct}, $I_{\alpha}$ converges within a few $e$-folds after horizon exit, after which this contribution depends on the wave vector only through $\mu$.
Since $\beta$ keeps growing during the anisotropic stage, $\mu$ changes at fixed comoving wavenumber of a generic oblique direction, $k_{x}k_{a}\neq0$, so that the contribution continues to evolve through $\beta(N)$ and becomes constant only after the background isotropizes and $\beta$ freezes.
For wave vectors along the axes, $\mu$ does not change, and Eq.~\eqref{eq:RhatNmu} approaches a constant even while the background remains anisotropic. For $k_{x}=0$, for instance, it reduces to $-c^{N}(\beta'+3d_{\beta}I_{\alpha})/(2-\beta')$.
Equation~\eqref{eq:RhatNmu} applies for $k_{a}\neq0$, since the constraints impose additional relations for $k_{a}=0$.
The anisotropy of the curvature perturbation at $N_{e}$ is therefore not determined by the anisotropy of the initial conditions alone.

If, after $N_{e}$, the background vector field becomes negligible and the expansion isotropizes, i.e., $\beta'\rightarrow0$ and $\mathcal{V}\rightarrow0$, the two conditions $J_{x}=0$ and $J_{a}=0$ coincide.
Moreover, $I_{\alpha}$ and $I_{\beta}$ converge, $D$ and the charges are frozen, and $\beta$ becomes constant, so that it can be absorbed into a rescaling of the spatial coordinates and $\Lop\rightarrow\nabla^{2}$.
Both amplitudes then reduce to
\begin{align}
    &\mathcal{R}_{C} = -c^{N} - c^{p}\qty(\frac{\partial_{\lambda^{p}}\phi}{\phi'} - W_{p}I_{\alpha}) + \Delta\mathcal{R}_{\rm ani}, \qquad R_{C}=-\frac{4}{a^{2}}\nabla^{2}\mathcal{R}_{C}, \label{eq:Rc}
\end{align}
where $\Delta\mathcal{R}_{\rm ani}$ is now a constant that depends on the direction of $\vec{k}$ relative to the $x$-axis.
Explicitly, with $\beta'\rightarrow0$, Eq.~\eqref{eq:DRani} becomes
\begin{align}
\Delta\mathcal{R}_{\rm ani}=P_{2}(\mu)\qty(D-\delta d_{\beta}I_{\alpha})-\mu^{2}\qty[e^{-6\beta}\mathcal{I}_{\Delta}\qty(W-\delta d_{\beta})+\frac{1}{2}\qty(1+3d_{\beta}e^{-6\beta}\qty(\mathcal{I}_{\Delta}-e^{6\beta}I_{\alpha}))\frac{\delta d_{A}}{d_{A}}], \label{eq:DRanimu}
\end{align}
where all quantities take their values frozen after isotropization.
The anisotropic stage thus leaves a quadrupolar term and a term proportional to $\mu^{2}$, the latter arising from the conserved quantities.
In the scalar-only case, Eq.~\eqref{eq:DRanimu} reduces to Eq.~\eqref{eq:DRiso}.
This has precisely the form of Eq.~(2.46) of Ref.~\cite{Sasaki:1998ug}, together with the relation ${}^{(3)}R=-(4/a^{2})\nabla^{2}\mathcal{R}$, up to the constant, direction-dependent offset $\Delta\mathcal{R}_{\rm ani}$ left by the anisotropic stage.
Whether this happens, however, depends on the physics after the anisotropic stage, which is beyond the scope of this paper.
For instance, when the coupling function becomes constant, Eq.~\eqref{eq:usof} gives $\bar{A}'_{x}\propto 1/(He^{\alpha})$, and Eq.~\eqref{eq:Acbeta} then implies $\beta'\propto e^{-\alpha}$ in a matter-dominated stage, whereas $\beta'$ approaches a constant in a radiation-dominated stage unless the electric field is screened.
We therefore regard Eqs.~\eqref{eq:RCx} and \eqref{eq:RCa} at $N=N_{e}$ as the output of our superhorizon analysis, and Eq.~\eqref{eq:Rc} as its form under the additional assumption of isotropization.
In either case, the curvature perturbation retains the memory of the anisotropic stage through the initial-condition parameters $(c^{N}, c^p)$.

\subsection{Relation to a tensor metric perturbation}
\label{GW}

In the preceding subsections, we evaluated the spatial curvature on various slicings using the conserved quantity $W$ and the superhorizon 
solutions derived in Sec.~\ref{LWP}. 
However, in anisotropic backgrounds with broken rotational symmetry, the perturbations that would be scalar and tensor modes in the isotropic case are mixed on superhorizon scales.
In this subsection, we define a gauge-invariant metric amplitude $G$ within the even-parity sector and derive its relation to the curvature perturbation. It reduces to the plus-polarized tensor mode in the isotropic limit.

In isotropic cosmology, tensor modes are gauge-invariant and defined as the transverse-traceless part of the metric perturbation.
In our anisotropic background, the reduced rotational symmetry prevents such a definition, and the curvature perturbation is mixed with the metric perturbation that plays the role of the tensor mode.

Following the metric parametrization of Ref.~\cite{Watanabe:2010fh}, we choose a scalar-flat slicing and a spatial gauge in which the even-parity spatial-metric perturbation takes the diagonal form below.
In this gauge, the even-parity metric perturbations take the form
\begin{align}
    \delta g_{\mu\nu}=e^{2\alpha}
\begin{pmatrix}
    -2e^{-2\alpha}H^{-2}\tilde{A} & e^{-4\beta}H^{-1}\tilde{B}_{x} & e^{2\beta}H^{-1}\tilde{B} & 0 \\
    \ast & e^{-4\beta}G & 0 & 0 \\
    \ast & \ast & e^{2\beta}G & 0 \\
    \ast & \ast & \ast & -e^{2\beta}G \\ \label{eq:G}
    \end{pmatrix}
    .
\end{align}
Here, $\ast$ denotes the symmetric part.
The amplitude $G$ obtained in Eq.~\eqref{eq:G} is gauge-invariant under the remaining gauge freedoms and
reduces to the plus mode of the gravitational wave in the isotropic limit.
More precisely, in the isotropic limit with $k_{x}\neq0$, the tensor $\mathrm{diag}(G,G,-G)$ is not transverse-traceless.
It is, however, related by a time-independent spatial gauge transformation to the transverse-traceless tensor of plus polarization defined with respect to $\vec{k}$ and the $z$-axis, whose amplitude equals $G$.\footnote{See also Ref.~\cite{Domenech:2025ccu} for the recent discussion on the gauge-invariant definition of gravitational wave strain.}
In the following, we obtain $G$ by transforming the spatial metric from the $\mathcal{N}$-gauge to this representation.

In this subsection we keep $C$ in the expressions, because the transformation to the gauge~\eqref{eq:G} involves spatial transformations that shift $C$. In the gauge~\eqref{eq:C0}, $C$ is set to zero in the final results.

In the $\mathcal{N}$-gauge and in Fourier space, we have
\begin{align}
    \delta g_{\mu\nu}=e^{2\alpha}
\begin{pmatrix}
    -2e^{-2\alpha}H^{-2}A & 0 & 0 & 0 \\
    \ast & e^{-4\beta}(2C-4D) & ie^{2\beta}k_yE & 0 \\
    \ast & \ast & e^{2\beta}\qty(2C+2D-k^2_{y}f) & 0 \\
    \ast & \ast & \ast & e^{2\beta}\qty(2C+2D+k^2_{y}f) \\ \label{eq:Ng}
    \end{pmatrix}
    ,
\end{align}   
where we have taken the wave vector to be $(k_x, k_y, 0)$ without loss of generality.
Since there is rotational freedom in the $(y,z)$-plane, we can take the direction arbitrarily just by taking the wave vector as $k^2_{y}\rightarrow k^2_{a}$, and we will take this rotation in the following.

To move from the $\mathcal{N}$-gauge to the flat slicing, we first perform a time shift $dN_{F}$~\eqref{eq:dNF}.
Additionally, the transformed perturbation must satisfy $g_{xy}=g_{yz}=0$ and $e^{4\beta}\delta g_{xx}=e^{-2\beta}\delta g_{yy}=-e^{-2\beta}\delta g_{zz}$. These equalities concern the perturbations $\delta g_{ij}$, not the full spatial metric.
After a straightforward calculation, the transformed lapse and shift perturbations in Eq.~\eqref{eq:G} are given by
\begin{align}
& \tilde{A} = A - dN'_{F} + \frac{H'}{H}dN_F, \quad \tilde{B}_{x} = - HL'^{x} + ik_{x}e^{-2\alpha+4\beta}H^{-1}dN_{F}, \quad \tilde{B} = -HL' + e^{-2\alpha-2\beta}H^{-1}dN_{F}.
\end{align}

The amplitude $G$ is obtained by imposing the conditions on the transformed spatial metric. 
For $\vec{k}=(k_x,k_y,0)$, $g_{yz}=0$ already holds in the even-parity sector. 
The two independent conditions determining $L^x$ and $L$ are $g_{xy}=0$ and $\delta g_{yy}+\delta g_{zz}=0$; the remaining diagonal relation follows from the vanishing intrinsic three-Ricci scalar for $k_a\neq0$. 
We find
\begin{align}
    G
    &= -2C-2D-k^2_{a}f + 2\qty(1+\beta')dN_{F} \nonumber \\
   &= -2\qty(D + C + \frac{k^2_{a}}{2}f) - \frac{e^{2\alpha}}{2}\qty(1+\beta')\qty(\Lopk +\beta'\LopDk)^{-1}R_{\mathcal{N}}. 
   \label{eq:G2} 
    \end{align}

The explicit forms of the spatial gauge transformation parameters that lead to these expressions are

\begin{align}
L^{x}_{k}
&=\frac{e^{6\beta}}{k^2_{a}}\qty[2ik_{x}(C+D) + k^2_{a}E -2i(1+\beta')k_{x}dN_{F,k}],    
\quad L_{k}
=-\frac{2[C + D -(1+\beta')dN_{F,k}]}{k^2_{a}}, 
    \end{align}
where we have assumed $k_a \neq 0$.
For exactly parallel wave vectors ($k_a=0$), the polarization basis used to define $G$ becomes degenerate; the axisymmetric even-parity perturbations then carry no helicity-two contribution in this sector. 
The expression obtained for generic $k_a\neq0$ should not be used to define $G$ in that special direction.
We emphasize that $G$ is invariant under the residual gauge transformations left unfixed by our gauge conditions.
The odd-parity sector of App.~\ref{VP} contains the metric perturbation that reduces to the cross-polarized tensor mode in the isotropic limit. 
It couples to the gauge-field perturbation $\delta A_a$ but does not mix with the even-parity curvature perturbation at linear order. The relation below therefore concerns the even-parity amplitude $G$.

Using $C_{F}=C-dN_{F}=-\mathcal{R}_{\mathcal{N}}$ and $D_{F}=D-\beta'dN_{F}$ defined in Sec.~\ref{sec:flat}, Eq.~\eqref{eq:G2} can be written as
\begin{align}
    G = 2\qty(\mathcal{R}_{\mathcal{N}}-D_{F})-k^{2}_{a}f, \label{eq:GF}
\end{align}
which expresses $G$ in terms of $\mathcal{R}_{\mathcal{N}}$, $D_{F}$ and $f$.
In the gauge fixed by Eqs.~\eqref{eq:LLx} and \eqref{eq:C0}, $f=d_{f}I_{\alpha}$, and Eqs.~\eqref{eq:dtheta} and \eqref{eq:df} give, in Fourier space,
\begin{align}
k^{2}_{a}d_{f} = 2\qty(2W+\delta d_{\beta}) + e^{6\beta(N_{0})}\frac{k^{2}_{x}}{k^{2}_{a}}\qty(-4W+4\delta d_{\beta}-6d_{\beta}\frac{\delta d_{A}}{d_{A}}).
\end{align}
Hence
\begin{align}
    G = 2\qty(\mathcal{R}_{\mathcal{N}}-D_{F}) - \qty[2\qty(2W+\delta d_{\beta}) + e^{6\beta(N_{0})}\frac{k^{2}_{x}}{k^{2}_{a}}\qty(-4W+4\delta d_{\beta}-6d_{\beta}\frac{\delta d_{A}}{d_{A}})]I_{\alpha}. \label{eq:GW}
\end{align}
The conserved quantities thus contribute to $G$ through the same memory integral $I_{\alpha}$ that appears in the curvature~\eqref{eq:RW}, while $\mathcal{R}_{\mathcal{N}}$ and $D_{F}$ carry the remaining Class I contribution.
For regular perturbations, the physical component $k^{2}_{a}d_{f}$ remains finite as $k_{a}\rightarrow0$, which requires $-4W+4\delta d_{\beta}-6d_{\beta}\delta d_{A}/d_{A}=O(k^{2}_{a})$. The term proportional to $k^{2}_{x}/k^{2}_{a}$ in Eq.~\eqref{eq:GW} then stays finite.
Since the polarization basis defined with respect to the direction of $k_{a}$ degenerates at $k_{a}=0$, the value of $G$ for exactly parallel modes should not be identified with this limit.

Equation~\eqref{eq:G2} reveals two important features.
First, $D+C+\frac{k_a^2}{2}f$ is invariant under spatial gauge transformations, and therefore characterizes the anisotropy of the spatial-metric perturbation in a spatial-gauge-invariant manner. Second, the amplitude $G$ depends on both the metric anisotropy $D$ and the intrinsic curvature $R_{\mathcal N}$.
Since the time-translation gauge mode has been removed by Eq.~\eqref{eq:C0}, all six modes $(c^{N}, c^{p})$ are physical. In particular, the dilatation mode $c^{N}$ contributes to $G$ through $D_{F}$ and $\mathcal{R}_{\mathcal{N}}$, given by Eq.~\eqref{eq:RhatN}, and through $\delta d_{\beta}$ and $\delta d_{A}$ in Eq.~\eqref{eq:GW}.
This establishes a common superhorizon transfer from the even-parity initial-condition parameters to the curvature perturbation and $G$. Their cross-correlation can be evaluated once the covariance of those initial data is specified.

\section{Conclusion}
\label{Con}

In this paper, we have studied the long-wavelength evolution of linear perturbations around an anisotropic Bianchi type I spacetime sourced by a scalar field coupled to a U(1) gauge field. Working within the spatial gradient expansion in the uniform-$\mathcal{N}$ gauge, we constructed the complete set of leading-order superhorizon solutions for generic wave vectors ($k_a\neq 0$). The even-parity perturbations with direct counterparts in the axisymmetric background are obtained from variations of its parameters, while the remaining even-parity and odd-parity perturbations can be understood as non-axisymmetric deformations of general Bianchi type I solutions and residual gauge transformations. This extends the background--perturbation correspondence of Ref.~\cite{Sasaki:1998ug} to anisotropic spacetimes. The Bianchi identities and Maxwell equations ensure that the momentum and Gauss constraints are preserved during evolution, so that they need only be imposed on the initial data, where they reduce to algebraic relations among the integration constants (App.~\ref{DGE}).

After fixing the residual gauge freedom, the solutions contain six physical parameters $(c^{N},c^{p})$ in the even-parity sector and four in the odd-parity sector, corresponding to five propagating degrees of freedom. We derived a conserved quantity $W$ that generalizes the Wronskian of Ref.~\cite{Sasaki:1998ug}. Unlike in the isotropic case, however, $W$ alone is insufficient to determine the curvature perturbation on uniform-$\mathcal{N}$ hypersurfaces. Together with the variations of the shear and electric-flux integration constants, $\delta d_{\beta}$ and $\delta d_A$, and the dynamical metric shear $D$, it allows this intrinsic curvature to be evaluated explicitly in terms of background variations and time integrals. 
The conserved comoving curvature perturbations are then obtained by combining this intrinsic-curvature contribution with the familiar $\delta N$ contribution arising from the change of time slicing.

An important consequence concerns the dilatation mode $c^{N}$, which corresponds to the adiabatic growing mode of Ref.~\cite{Sasaki:1998ug} in the isotropic limit. 
Although its Wronskian vanishes, this mode carries nonzero variations of the background charges, $\delta d_{\beta}=-3c^{N}d_{\beta}$ and $\delta d_A=-2c^{N}d_A$. 
Consequently, the comoving curvature perturbation can evolve even when only this mode is excited. 
This mode-by-mode statement refers to the residual spatial gauge fixed in Sec.~\ref{SS}, whereas the total comoving curvature perturbation is independent of this choice. For attractor-like backgrounds with nearly constant $H$ and small shear, the memory integrals approach constant values within a few $e$-folds after horizon exit. 
Nevertheless, their contributions to the curvature can continue to evolve through the changing physical direction of the wave vector during anisotropic expansion. 

We evaluated the two directional comoving curvature perturbations at the end of the anisotropic stage in terms of superhorizon initial data through Eq.~\eqref{eq:ci}. 
In the scalar-only case with the background shear switched off, the result reduces to that of Ref.~\cite{Sasaki:1998ug}, apart from the direction-dependent constant contribution associated with the anisotropic spatial rescaling $c^{\beta_0}$, which contains a long-wavelength tensor component. The additional contribution $\Delta\mathcal{R}_{\rm ani}$ encodes the history of anisotropic expansion and may leave imprints in the statistical anisotropy of primordial perturbations.

We also derived an explicit relation between the curvature perturbation and the metric perturbation $G$, which reduces to the plus-polarized tensor mode in the isotropic limit.  
The conserved quantities enter $G$ explicitly through the memory integral $I_\alpha$, which also contributes to the curvature, making their common superhorizon transfer structure explicit.

We close by distinguishing the results expected to hold generally from those specific to the present setup. 
The correspondence between long-wavelength perturbations and variations of homogeneous background solutions, together with the role of the constraint equations, reflects the general separate-universe structure at leading order in the gradient expansion. 
Likewise, global coordinate and U(1) transformations that are pure gauge in the homogeneous limit can generate nontrivial long-wavelength perturbations when their parameters vary spatially. 
The integration constants of the traceless Einstein equations are also related to the shear Noether charges identified in Ref.~\cite{Tanaka:2021dww}. 
These structures are consistent with the general anisotropic separate-universe framework of Ref.~\cite{Tanaka:2021dww}, while their explicit realization in terms of the conserved quantities and perturbation solutions is specific to the present model.
In contrast, the even--odd decomposition and the split of the ten physical parameters into six even-parity and four odd-parity ones rely on the axisymmetry of the background. 
The specific form of $W$, the shear and electric-flux contributions to the curvature, and the relation~\eqref{eq:GW} depend on the matter content considered here. 
For other matter content or background configurations, such as charged matter or a magnetic background, the relevant conserved quantities and transfer relations may differ.

Several important directions remain for future investigation. 
Connecting the superhorizon solutions obtained here to quantum fluctuations generated around horizon crossing will enable quantitative predictions for primordial power spectra and their statistical anisotropy. 
This requires matching the long-wavelength solutions to finite-wavenumber mode functions, which is beyond the scope of the present analysis.
A detailed comparison with previous studies~\cite{Gumrukcuoglu:2010yc,Watanabe:2010fh,Bartolo:2012sd,Naruko:2014bxa,Fujita:2017lfu} will be presented in a separate publication. 
The present framework also provides a basis for computing curvature--tensor cross-correlations and higher-order correlation functions. 
Finally, the nonlinear extension discussed in App.~\ref{NL} offers a starting point for investigating non-Gaussianity in anisotropic cosmologies. 
This extension is restricted to the growing-mode sector and neglects the decaying shear components after the initial time; it therefore does not by itself reproduce all the memory contributions retained in the linear analysis.

\section*{Acknowledgments}

D.S. would like to thank Danilo Artigas for the fruitful discussion.
A.N. would like to thank Eiichiro Komatsu and Masahide Yamaguchi for their invaluable discussions and collaborative efforts during the early stages of this research.
The authors would also like to thank Takahiro Tanaka for the fruitful discussion.
D.S. thanks the Atom research fellow
program at YITP, Kyoto University for kind hospitality during part of this research.
This work of D.S. was partially supported by JSPS KAKENHI Grant Number JP24KJ1223, and the National Research Foundation of Korea Grant funded by the Korean Government RS-2024-00336507.
The work of A.N. was partly supported by JSPS KAKENHI Grant Numbers JP20H05852, JP23H01171, JP23K25868 and 26H02044.

\appendix

\section{Linearized perturbation equations}
\label{PEspacetime}

\subsection{Basic equations}

We collect the linearized component equations before the even--odd decomposition, keeping $C$ arbitrary in the $\mathcal{N}$-gauge. In the main text, the residual gauge condition $C=0$ is imposed from the end of Sec.~\ref{SS} onward.
At the linear order in the perturbations, the components of the energy-momentum tensor are given as
\begin{align}
    &\delta\rho = H^2\phi'\delta\phi'-H^2\phi'^2A+\pt{U}{\phi}\delta\phi+\frac{1}{2}\delta\mathcal{V},\\
    &\delta P=H^2\phi'\delta\phi'-H^2\phi'^2A-\pt{U}{\phi}\delta\phi+\frac{1}{6}\delta\mathcal{V}, \\
    &J_{i}=- H\phi'\partial_{i}(\delta\phi)-\frac{d_{A}}{e^{3\alpha}}\qty(\partial_{a}\delta A_{x}- \partial_{x}\delta A_{a})\delta^{a}_{\ i}, \label{eq:AJi} \\
    &\delta\pi^{x}_{\ x}=-\frac{2}{3}\delta\mathcal{V}, \label{eq:defpixx} \\
    &\delta\pi^{x}_{\ a}=-H^2I^2e^{-2\alpha+4\beta}\bar{A}'_{x}\delta A'_{a}, \\
    &\delta\pi^{a}_{\ b}=\frac{1}{3}\delta\mathcal{V}\delta^{a}_{\ b},
\end{align}
where $\delta\mathcal{V}$ is given in Eq.~\eqref{eq:defdV}. At leading order in the gradient expansion, the terms with spatial derivatives in the following equations are dropped.

The scalar field equation is given by
 \begin{align}
 &H\partial_{N}(H\delta\phi')
 +3H^2\delta\phi'-2AH\partial_{N}(H\phi')-
 H^2\qty(A'+6A)\phi'+\ptt{U}{\phi}\delta\phi - e^{-2\alpha-2\beta}\qty(e^{6\beta}\partial^2_{x} + \partial^2_{a})\delta\phi \nonumber \\
 &= \frac{\partial_{\phi}I}{I}\delta\mathcal{V}+\frac{\partial^2_{\phi}I}{I}\mathcal{V}\delta\phi-\qty(\frac{\partial_{\phi}I}{I})^2\mathcal{V}\delta\phi. \label{eq:ApertphiG}
 \end{align}

 For the vector field, the spatial components of the equations of motion are expressed as
 \begin{align}  
 &H\partial_{N}\qty[I^2e^{\alpha+4\beta}H\bar{A}'_{x}\qty(\frac{\delta A'_{x}}{\bar{A}'_{x}}+2\frac{\partial_{\phi}I}{I}\delta\phi-A+C+4D)]  - e^{-\alpha+2\beta}I^2 \qty(\partial^2_{a}\delta A_{x} - \partial_{x}\partial_{a}\delta A_{a}) = 0, \label{eq:PMaxx} \\
 &H\partial_{N}\qty[I^2e^{\alpha+4\beta}H\bar{A}'_{x}\qty(e^{-6\beta}\frac{\delta A'_{a}}{\bar{A}'_{x}}-E_{a})] + e^{-\alpha+2\beta}I^2 \qty(\partial_{a}\partial_{x}\delta A_{x} - \partial^2_{x}\delta A_{a}) = 0, \label{eq:PMaxa}
      \end{align}
and the time component gives the constraint
 \begin{align}  
 &\partial_{x}\qty(\frac{\delta A'_{x}}{\bar{A}'_{x}}+2\frac{\partial_{\phi}I}{I}\delta\phi-A+C+4D)+ \partial_{a}\qty(e^{-6\beta}\frac{\delta A'_{a}}{\bar{A}'_{x}}-E_{a}) = 0. \label{eq:PMaxt}
      \end{align}

The Hamiltonian constraint can be written as
\begin{align}
    &6H^2\qty[(1-\beta'^2)A+\beta'D'] + e^{-2\alpha -2\beta}\qty[(2e^{6\beta}\partial^2_{x}+2\partial^2_{a})C + (2e^{6\beta}\partial^2_{x}-\partial^2_{a})D - e^{6\beta}\partial_{x}\partial_{a}E_{a} - \partial_{a}\partial_{b}f_{ab}] \nonumber 
\\
    & =-\delta\rho.
    \label{eq:pert00G}
\end{align}
The momentum constraint for $x$-direction reads
\begin{align}
&-2H(1+\beta')\partial_{x}A+6H\beta'\partial_{x}(C+D)+2H\partial_{x}D'-\frac{H}{2}\qty(\partial_{a}E'_{a}+6\beta'\partial_{a}E_{a})=J_{x}, \label{eq:pert0x} 
\end{align}
and $a$-direction
\begin{align}
    &H\qty[-(2-\beta')A-3\beta'(C-2D)-D']_{,a}-   
    H\partial_{b}f'_{ab}-\frac{H}{2}e^{6\beta}\partial_{x}E'_{a}=J_{a}. \label{eq:pert0a} 
\end{align}
For the spatial components, the trace part is given as 
\begin{align}
    &2H^2A'-6
    H^2\beta'D'+2\qty[2HH'+3
    H^2\qty(1+\beta'^2)]A  \nonumber \\
    &+ \frac{e^{-2\alpha -2\beta}}{3}\qty[2(e^{6\beta}\partial^2_{x}+\partial^2_{a})A+ (2e^{6\beta}\partial^2_{x}+2\partial^2_{a})C + (2e^{6\beta}\partial^2_{x}-\partial^2_{a})D - e^{6\beta}\partial_{x}\partial_{a}E_{a} - \partial_{a}\partial_{b}f_{ab}] =\delta P, \label{eq:perttrG}
\end{align}
and the traceless components are written as    
\begin{align}
    &-2\qty[H\partial_{N}(HD')+3H^2D'-2\qty(H\partial_{N}(H\beta')+3H^2\beta')A-H^2\beta'A'] \nonumber \\
     &+ \frac{e^{-2\alpha -2\beta}}{3}\qty[(-2e^{6\beta}\partial^2_{x}+\partial^2_{a})A+ (-2e^{6\beta}\partial^2_{x}+\partial^2_{a})C -2 (e^{6\beta}\partial^2_{x}-2\partial^2_{a})D + e^{6\beta}\partial_{x}\partial_{a}E_{a} - 2\partial_{a}\partial_{b}f_{ab}]=\delta\pi^{x}_{\ x}, \label{eq:pertxx} \\
    &\frac{e^{6\beta}}{2}\qty[H\partial_{N}(HE'_{a})+H^2\qty(3+6\beta')E'_{a}] + \frac{e^{-2\alpha +4\beta}}{2}\qty[- 2\partial_{x}\partial_{a}A - 2\partial_{x}\partial_{a}C - 2\partial_{x}\partial_{a}D +  2\partial_{x}\partial_{b}f_{ba}] =\delta\pi^{x}_{\ a}, \label{eq:pertxa} \\
&H\partial_{N}(Hf'_{yy})+3H^2{f}'_{yy} \nonumber \\
&+ \frac{e^{-2\alpha -2\beta}}{2}\qty[-(\partial^2_{y}-\partial^2_{z})A -(\partial^2_{y}-\partial^2_{z})C+2(\partial^2_{y}-\partial^2_{z})D + e^{6\beta}\partial_{x}(\partial_{y}E_{y}-\partial_{z}E_{z}) -2 e^{6\beta}\partial^2_{x}f_{yy}]=0,  \label{eq:pertyy} \\
&H\partial_{N}(Hf'_{yz})+3H^2{f}'_{yz} + \frac{e^{-2\alpha -2\beta}}{2}\qty[-2\partial_{y}\partial_{z}A -2\partial_{y}\partial_{z}C +4\partial_{y}\partial_{z}D + e^{6\beta}\partial_{x}(\partial_{y}E_{z}+\partial_{z}E_{y}) -2 e^{6\beta}\partial^2_{x}f_{yz}]=0.  \label{eq:pertyz} 
    \end{align}

\subsection{Spatial gradient expansion}
\label{DGE}

In this appendix, we summarize the order-by-order structure of the equations used in the gradient expansion and derive the consistency relations among them.

As in Sec.~\ref{PE}, we expand in powers of $\epsilon$, assigning one power of $\epsilon$ to each spatial derivative, and write the regular perturbation components as
\begin{align}
    \mathcal{P} & = \sum_{n=0}^{\infty}\mathcal{P}^{(n)} \epsilon^n, \quad \mathcal{P}:=\qty{A, C, D, E_{a}, f_{ab}, \delta\phi, \delta A_{x}, \delta A_{a}}.
    \label{eq:GE}
\end{align}
Here $\epsilon$ serves as a formal expansion parameter that tracks the order in spatial gradients.
    
The order-by-order equations follow from App.~\ref{PEspacetime} by counting each spatial derivative as one higher order in $\epsilon$. 
Thus, terms with two spatial derivatives in the evolution equations and the Hamiltonian constraint involve $\mathcal{P}^{(n-2)}$ at order $n$, while the one-gradient momentum and Gauss constraints involve $\mathcal{P}^{(n-1)}$, with $\mathcal{P}^{(n)}=0$ for $n<0$.

We denote the Klein--Gordon, Maxwell, and Einstein equations at order $n$ by $\mathcal{F}^{(n)}=0$, $\mathcal{M}_{\mu}{}^{(n)}=0$ and $\mathcal{E}^{\mu}{}_{\nu}{}^{(n)}=0$, respectively.

By direct computation, we find that the relation between the equations
\begin{align}
    &3H\qty(\mathcal{E}^t{}_t{}^{(n)} 
    - \frac{1}{3}\mathcal{E}^i{}_i{}^{(n)} ) 
    + 3H\beta'\qty(\mathcal{E}^x{}_x{}^{(n)}
    -\frac{1}{3}\mathcal{E}^i{}_i{}^{(n)}) 
    - H\phi' \mathcal{F}^{(n)}
    + H\partial_{N}\mathcal{E}^t{}_t{}^{(n)} 
    + \frac{d_{A}}{e^{3\alpha}I^2}\mathcal{M}_x{}^{(n)} =  
    -\partial_{i}\mathcal{E}^i{}_t{}^{(n-1)}, \\
        &3H\mathcal{E}^t{}_i{}^{(n)}  + H\partial_{N}\mathcal{E}^t{}_i{}^{(n)}   
    +  \frac{d_{A}}{I^2e^{\alpha+4\beta}}\mathcal{M}_t{}^{(n)}\delta^x_i =  
    -\partial_{j}\mathcal{E}^j{}_i{}^{(n-1)}
    \label{eq:ConRel}
\end{align}
holds for $n \geq 0$.
These identities follow from the contracted Bianchi identities and the matter-field equations, with $\mathcal{E}^{\mu}{}_{\nu}$ understood as the mixed-index Einstein-equation residual. In particular, the spatial divergence in the first identity carries the indices $\mathcal{E}^{i}{}_{t}$.
The identities determine how the constraints propagate once the Hamiltonian constraint and the leading-order evolution and matter-field equations are satisfied. At the first nontrivial gradient order, $n=1$, spatial derivatives of the zeroth-order equation residuals vanish, so the momentum constraints obey homogeneous propagation equations. The Maxwell identity $\nabla_{\mu}\mathcal{M}^{\mu}=0$ similarly gives homogeneous evolution for the Gauss constraint. Therefore, it is sufficient to impose the momentum and Gauss constraints on the initial data, where they become algebraic relations among the integration constants of the leading-order superhorizon solutions.

\subsection{Varied background equations}
\label{app:LO}

Varying the background scalar-field and Einstein equations, Eqs.~\eqref{eq:ABGphi}, \eqref{eq:ABG00}, \eqref{eq:ABGtr}, and \eqref{eq:Abeta1eq}, yields the following equations. The variation of the background Maxwell equation is given separately in Eq.~\eqref{eq:bpertmax}.
\begin{align}
 &H\partial_{N}\qty(H\Delta\phi')+2\Delta H\partial_{N}\qty(H\phi')+H^2\qty{\qty(\frac{\Delta H}{H})'+6\frac{\Delta H}{H}}\phi'+3H^2(\Delta\phi)'+\ptt{U}{\phi}\Delta\phi \nonumber \\
    &=\frac{\partial_{\phi}I}{I}\Delta\mathcal{V}+\frac{\partial^2_{\phi}I}{I}\mathcal{V}\Delta\phi-\qty(\frac{\partial_{\phi}I}{I})^2\mathcal{V}\Delta\phi, \label{eq:bpertscalar} \\
    &6H^2\qty[-(1- \beta'^2)\frac{\Delta H}{H}+\beta'(\Delta \beta)']=-\Delta\rho, \label{eq:bpert00} \\
    &2H^2\qty[\qty(\frac{\Delta H}{H})'+\qty(3+3\beta'^{2}+2\frac{H'}{H})\frac{\Delta H}{H}+3\beta'(\Delta\beta)']=-\Delta P, \label{eq:bperttr} \\    
&H\partial_{N}(H\Delta \beta')+3H^2\qty(\Delta\beta)'+2\qty[H\partial_{N}(H\beta')+3H^2\beta']\frac{\Delta H}{H}+H^2\beta'\qty(\frac{\Delta H}{H})'=\frac{1}{3}\Delta\mathcal{V}, \label{eq:bpertxx} 
\end{align}
where we have introduced
\begin{align}
&\Delta\mathcal{V}
     =2\mathcal{V}\qty(\frac{\Delta \bar{A}'_{x}}{\bar{A}'_{x}}+\frac{\partial_{\phi}I}{I}\Delta\phi+\frac{\Delta H}{H}-\Delta\alpha+2\Delta\beta), \\
    &\Delta\rho=H^2\phi'\Delta\phi'+H\Delta H\phi'^{2}+\pt{U}{\phi}\Delta\phi+\frac{1}{2}\Delta\mathcal{V}, \\
    &\Delta P=H^2\phi'\Delta\phi'+H\Delta H\phi'^{2}-\pt{U}{\phi}\Delta\phi+\frac{1}{6}\Delta\mathcal{V}.
\end{align}

\subsection{Relation between the two parametrizations of $E$}
\label{app:dE}

Integrating Eq.~\eqref{eq:EpSol} once more, we obtain
\begin{align}
&E=\tilde{c}_{E}+\int_{N_0}^{N}e^{-3\alpha(N_*)-6\beta(N_*)}\qty(\widetilde{d}_{E}-2d_{A}\delta A(N_*))\frac{dN_{*}}{H(N_*)}, \label{eq:Esol}
\end{align}
where $\widetilde{c}_E$ is an integration constant.
We note that $\widetilde{c}_E$ and $\widetilde{d}_E$ are related to the
$c_E$ and $d_E$ by a linear transformation.
The explicit relations between them are given by
\begin{align}
\widetilde{c}_{E}&=e^{-6\beta(N_{0})}c_{E} - \frac{d_{\perp}}{d_{A}}, \\
\widetilde{d}_E & =2d_{A}\delta A -6H\beta'e^{3\alpha}c_{E} +  d_{E} \qty(e^{6\beta}  -6H\beta' e^{3\alpha} I_{\beta}), \label{eq:dEdE}
\end{align}
where $N_{0}$ is the initial time for the superhorizon evolution, and we have fixed the lower limit of the integration $I_{\beta}$ as $N_0$.

Using the simplified expressions obtained after the residual gauge transformation in Sec.~\ref{SS} and Eq.~\eqref{eq:dEdE}, we have
\begin{align}
\widetilde{d}_E & =d_{E} \qty(e^{6\beta}  -6H\beta' e^{3\alpha} I_{\beta}(N) + 2d_{A}I_{A\beta}(N)). 
\end{align}
The quantity in parentheses has zero $N$ derivative by $\bigl(He^{3\alpha}\beta'\bigr)'=d_A\bar A'_x/3$, $I_\beta'=e^{-3\alpha+6\beta}/H$, and $I_{A\beta}'=\bar A'_x I_\beta$. With both integrals vanishing at $N_0$, evaluation there gives Eq.~\eqref{eq:dEtilde}.

\section{Conservation and Wronskian representation of $W$}
\label{app:Wronskian}

\subsection{Conservation of $W$}
\label{app:Wcons}

We demonstrate the conservation law $W'=0$ at leading order in the spatial gradient expansion.
Differentiating Eq.~\eqref{eq:Wdef} and using that $d_{A}$ and $\delta d_{A}$ are constants, we have
\begin{align}
W' &= He^{3\alpha}\qty[\qty(3+\frac{H'}{H})\qty(A-\frac{1}{2}\phi'\delta\phi-3\beta'D) + A' - \frac{1}{2}\qty(\phi''\delta\phi+\phi'\delta\phi') - 3\qty(\beta''D+\beta'D')] \nonumber\\
&\quad + \frac{1}{6}\qty(\delta d_{A}\,\bar{A}'_{x} - 2d_{A}\,\delta A'_{x}).
\end{align}
The trace part of the spatial Einstein equations~\eqref{eq:perttrG} gives
\begin{align}
A' = 3\beta'D' - \qty(2\frac{H'}{H}+3+3\beta'^{2})A + \frac{\delta P}{2H^{2}}, \qquad
\delta P = H^2\phi'\delta\phi' - H^2\phi'^2A - \partial_{\phi}U\delta\phi + \frac{1}{6}\delta\mathcal{V}.
\end{align}
Substituting this into $W'$, the terms with $D'$ and $\delta\phi'$ cancel.
We then use the background scalar-field equation~\eqref{eq:ABGphi} for $\phi''$, Eq.~\eqref{eq:Abeta1eq} for $\beta''$, and
$H'/H = -\phi'^{2}/2 - \mathcal{V}/(3H^{2}) - 3\beta'^{2}$, which follows from Eqs.~\eqref{eq:ABG00} and \eqref{eq:ABGtr}.
With Eqs.~\eqref{eq:defdV} and \eqref{eq:nu2}, all remaining terms are proportional to $\mathcal{V}$ and combine into
\begin{align}
W' = \frac{e^{3\alpha}\mathcal{V}}{6H}\qty[\frac{\delta d_{A}}{d_{A}} - \frac{\delta A'_{x}}{\bar{A}'_{x}} - 2\frac{\partial_{\phi}I}{I}\delta\phi + A - C - 4D],
\end{align}
The bracket vanishes by the integrated Maxwell equation~\eqref{eq:defCx}, establishing $W'=0$ without using either the perturbative Hamiltonian constraint or the perturbed scalar-field equation.

\subsection{Wronskian representation}

To see the structure of $W$, we first derive an auxiliary relation from the Hamiltonian constraint.
From Eqs.~\eqref{eq:ABG00} and \eqref{eq:pert00G}, we obtain
\begin{align}
 (2U+\mathcal{V})A=-H^2\phi'\delta\phi'-\pt{U}{\phi}\delta\phi-6H^2\beta'D'-\frac{1}{2}\delta\mathcal{V}.
 \label{eq:2UA}
\end{align}   
Using the scalar-field equation~\eqref{eq:ApertphiG} and the background equations~\eqref{eq:ABG00} and \eqref{eq:ABGtr} with $\gamma=0$, this can be rearranged into
\begin{align}
 H^2\qty(3+\frac{H'}{H})\qty(A-\frac{1}{2}\phi'\delta\phi-3\beta'D)=H^2\qty(3+\frac{H'}{H})w+\qty(\frac{1}{6}A+\frac{1}{2}C-2D-\frac{\pt{I}{\phi}}{I}\delta\phi-\frac{1}{2}\frac{\delta A'_{x}}{\bar{A}'_{x}})\mathcal{V}, \label{eq:W2}
\end{align} 
where the combination $w$ is defined as
\begin{align}
 w&:=\frac{1}{2}\qty(3+\frac{H'}{H})^{-1}\qty[\phi''\delta\phi-\phi'\delta\phi'+6\qty(\beta''D-\beta'D')].
\end{align} 
This $w$ is a Wronskian-like combination of the scalar-field and shear perturbations.
The representation in terms of $w$ assumes $3+H'/H\neq0$, as is the case during inflation.
Using the background-perturbation correspondence~\eqref{eq:delchi}, it can be rewritten as
\begin{align}
 w&=\frac{1}{2}\qty(3+\frac{H'}{H})^{-1}c^{J}\qty[\phi''\partial_{J}\phi-\phi'\partial_{J}\phi'+6\qty(\beta''\partial_{J}\beta-\beta'\partial_{J}\beta')].
\end{align} 
Substituting this expression into Eq.~\eqref{eq:W2} and using the definition of $W$~\eqref{eq:Wdef}, we obtain an explicit representation of $W$ in terms of the parameter shifts $c^J$ :
\begin{align}
 \qty(3+\frac{H'}{H})W
 &=He^{3\alpha}\qty(3+\frac{H'}{H})w \nonumber \\
 &+c^{J}\qty[\frac{e^{3\alpha}}{6H}\qty{-3+\qty(3+\frac{H'}{H})\frac{\bar{A}_{x}}{\bar{A}'_{x}}}\mathcal{V}\frac{\partial_{J}d_{A}}{d_{A}}+\frac{e^{3\alpha}}{3H}\qty{\frac{\partial_{J}H}{H}+3\partial_{J}\alpha-\qty(3+\frac{H'}{H})\frac{\partial_{J} \bar{A}_{x}}{\bar{A}'_{x}}
 }\mathcal{V}], \label{eq:Ww}
\end{align} 
where we have used Eq.~\eqref{eq:nu2}.
Unlike the usual matter-field Wronskian representation in isotropic multi-field inflation~\cite{Sasaki:1998ug}, the expression for $W$ in the present variables contains the metric perturbations $A$ and $D$, reflecting the dynamical shear and the anisotropic stress of the gauge field.

\section{Odd-parity perturbations}
\label{VP}

We now analyze the odd-parity perturbations defined in Sec.~\ref{Dec}, working with generic wave vectors for which the transverse direction in the $(y,z)$ plane is well defined.

We can write the perturbed metric as
\begin{align}
    ds^2&=-\frac{dN^2}{H^2}+2e^{2\alpha+2\beta}B_{a}H^{-1}dNdx^{a} \nonumber \\
    &+e^{2\alpha}\qty{e^{-4\beta}dx^2+2e^{2\beta}E_{a}dxdx^{a} +e^{2\beta}\qty(\partial_{a}f_{b}+\partial_{b}f_{a})dx^{a}dx^{b}},
    \end{align}
where $f_{a}$ satisfies  $\delta^{ab}\partial_{a}f_{b}=0$.
Note that the odd-parity part of $f_{ab}$ is normalized here as $f^{(\mathrm{o})}_{ab}=\frac{1}{2}(\partial_{a}f_{b}+\partial_{b}f_{a})$ in terms of the $2f_{ab}$ of Eq.~\eqref{eq:metp}; this factor is to be taken into account when the equations of App.~\ref{PEspacetime} are specialized to the odd-parity sector.

Under the coordinate transformation by $\delta x^{\mu}$, the metric components are transformed as
\begin{align}
    \delta g_{\mu\nu}\rightarrow \delta g_{\mu\nu} - \bar{g}_{\alpha\nu}\partial_{\mu}(\delta x^{\alpha}) - \bar{g}_{\mu\alpha}\partial_{\nu}(\delta x^{\alpha})-\partial_{\alpha}(\bar{ g}_{\mu\nu})\delta x^{\alpha}.
    \end{align}
For an odd-parity spatial generator $\delta x^\mu=(0,0,L^a)$ with $\partial_aL^a=0$, the metric perturbations transform as

\begin{align}
    &B_{a}\rightarrow B_{a}- HL'^{a}, \quad E_{a}\rightarrow E_{a} - \partial_{x}L^{a}, \quad f^{(\mathrm{o})}_{a}\rightarrow f^{(\mathrm{o})}_{a} - L^{a}.
    \end{align}

For the odd-parity perturbations, the relation
\begin{align}
\partial_{a}B_{a}=0 \label{eq:VNg1}
    \end{align}
is satisfied.
We work on uniform-$\mathcal{N}$ hypersurfaces and fix the odd-parity threading by imposing
\begin{align}
B_{a}=0. \label{eq:VNg2}
    \end{align}

Using the residual rotation symmetry, we take $\vec{k}=(k_x,k_y,0)$ and assume $k_y\neq0$ in the following; the exactly parallel case requires a separate decomposition.
Then, $E_{a}$ and $f_{a}$ can be written as
\begin{align}
E_{a}=(0, E_{z}), \quad f_{a}=(0, f_{z}),
    \end{align}
respectively.
The vector field is given by
\begin{align}
\delta A_{a}=\qty(0, \delta A_{z}).
    \end{align}

The odd-parity sector retains the residual gauge freedom
\begin{align}
    &z\rightarrow z + L^z({\bf x}). \label{eq:resgaugez}
    \end{align}
Under this transformation, the components of the spatial metric shift as
\begin{align}
    &E_z \rightarrow E_z - ik_xL^z({\bf x}), \quad f_z \rightarrow f_z - L^z({\bf x}). \label{eq:resgaugeEz}
    \end{align}
We note that there is no odd-parity counterpart of $L^x$.
Moreover, $\partial_a\theta$ is purely longitudinal, so the U(1) transformation does not act on $\delta A_z$ at all.
In particular, $E_z-ik_xf_z$ is invariant under the residual odd-parity spatial transformation~\eqref{eq:resgaugeEz}; a time-slicing change does not induce odd-parity spatial-metric components on the diagonal background.

The perturbed Einstein equations are written as
\begin{align}
    &H\partial_{N}(HE'_{z})+H^2(3+6\beta')E'_{z}=-2e^{-2\alpha-2\beta}H^2I^2\bar{A}'_{x}\delta A'_{z}, \label{eq:VpertE} \\
&H\partial_{N}(Hf'_{z})+3H^2f'_{z}=0 \label{eq:Vpertf} . 
    \end{align}

The Maxwell equations can be expressed as
\begin{align}
    0&=\frac{H}{e^{3\alpha}}\partial_{N}\qty[I^2e^{\alpha-2\beta}H\qty(\delta A'_{z}-e^{6\beta}\bar{A}'_{x}E_{z})], \label{eq:VMax} 
     \end{align}
and it is integrated to give
     \begin{align}
\delta A'_{z}&=\bar{A}'_{x}e^{6\beta}\qty(\frac{d^{z}_{\perp}}{d_{A}}+E_{z}), \label{eq:dAsol}
     \end{align}
with $d^{z}_{\perp}$ being a constant.

Equation~\eqref{eq:VpertE} can be written as
\begin{align}
    &\partial_{N}\qty(e^{3\alpha+6\beta}HE'_{z})=-2d_{A}\delta A'_{z},
    \end{align}

This can be integrated as
\begin{align}
E_z'=H^{-1}e^{-3\alpha-6\beta}(\widetilde d_E^z-2d_A\delta A_z),\nonumber\\
E_z=\widetilde c_E^z+\int_{N_0}^{N}\frac{e^{-3\alpha(N_*)-6\beta(N_*)}}{H(N_*)}\bigl(\widetilde d_E^z-2d_A\delta A_z(N_*)\bigr)\,dN_*.
\label{eq:Ezsol}
\end{align}
where $\tilde{c}^{z}_{E}$ and $\tilde{d}^{z}_{E}$ are constants for integration. 
As in the even-parity solution for $E$, we can express $E_z$ as
\begin{align}    
e^{6\beta}E_z &=  c^z_E -e^{6\beta}\frac{d^{z}_{\perp}}{d_{A}} + d^z_{E} I_{\beta}. \label{eq:Ezsol2}   
    \end{align}
Here, $c^{z}_{E}$ and $d^{z}_{E}$ are integration constants, and we obtain the following relations between the constants:
\begin{align}
\tilde{c}^z_{E}&=e^{-6\beta(N_{0})}c^z_{E} - \frac{d^{z}_{\perp}}{d_{A}}, \\
\widetilde{d}^z_E =2d_A\delta A_z-6H\beta'e^{3\alpha}c_E^z+d_E^z\bigl(e^{6\beta}-6H\beta'e^{3\alpha}I_\beta\bigr),\label{eq:dEz}
\end{align}

Using Eqs.~\eqref{eq:dAsol} and \eqref{eq:Ezsol2}, the perturbation of the vector field is expressed as
\begin{align}
    &\delta A_z' = \bar{A}'_{x} \qty(c^z_{E} + d^z_{E}I_{\beta}), 
         \end{align}
that is integrated to yield
 \begin{align}        
    &  \delta A_z =  c^z_{\delta A} + c^z_{E}\bar{A}_{x} + d^z_{E}I_{A\beta}(N).
        \label{eq:delAzsol} 
     \end{align}
Here, $c^z_{\delta A}$ is a constant of integration.

From Eq.~\eqref{eq:Vpertf}, we obtain
\begin{align}
f_z=c_f^z+d_f^z I_\alpha(N),\qquad I_\alpha(N)=\int_{N_0}^{N}\frac{dN_*}{H(N_*)e^{3\alpha(N_*)}}.\label{eq:fzsol}
\end{align}

In the Fourier space, the momentum constraint is given as
\begin{align}
 &ik^2_yf'_{z}+e^{6\beta}k_{x}E'_{z}=-2e^{-2\alpha+4\beta}I^2\bar{A}'_{x}k_{x}\delta A_{z}.
 \label{eq:VMC}
 \end{align}
 
By substituting Eqs.~\eqref{eq:Ezsol} and \eqref{eq:fzsol} into Eq.~\eqref{eq:VMC}, we obtain the relation between the constants $\tilde{d}_E^z$ and $d_f^z$:
 
\begin{align}
  \therefore &k_x\tilde{d}^{z}_{E}+ik^2_{y}d^{z}_{f}=0.
   \label{eq:MCz} 
    \end{align}

The integration constant $c^z_{f}$ can be eliminated by the residual gauge degrees of freedom.
To eliminate these unphysical degrees of freedom, we perform a residual gauge transformation~\eqref{eq:resgaugez} with
\begin{align}
 &L^z({\bf x}) = c^{z}_{f}. \label{eq:LLxV}
    \end{align}
Under this transformation, the solutions simplify to
\begin{align}    
E_z&= e^{-6\beta}c^z_E-\qty(\frac{d^{z}_{\perp}}{d_{A}} + ik_xc^z_{f}) + e^{-6\beta} d^{z}_{E}I_{\beta}(N), \quad f_z=d^z_{f}I_{\alpha}(N).
    \end{align}

After fixing the residual odd-parity spatial gauge, the independent data can be represented by $(d^z_{\perp}/d_A+ik_xc_f^z,c_E^z,d_E^z,c_{\delta A}^z,d_f^z)$. For $k_y\neq0$ and $d_A\neq0$, one momentum constraint~\eqref{eq:MCz} leaves four physical initial-data parameters, corresponding to two propagating degrees of freedom.
Evaluating Eqs.~\eqref{eq:dEz} and \eqref{eq:MCz}  at $N_0$, the constraint reads
\begin{align}
k_x \qty(2d_{A}c^z_{\delta A} -6d_{\beta}c^z_{E} +  d^z_{E} e^{6\beta(N_0)}) + i k^2_y d^z_f = 0, 
\end{align}
so that $d_f^z$ is determined by the other constants. In particular, $c_{\delta A}^z$ generates no electric field but a long-wavelength magnetic field.

The structure of these constants parallels that of the even-parity sector. The constant $\tilde d_E^z$ is the off-diagonal component $Q^x{}_z$ of the shear Noether charge in Eq.~\eqref{eq:NQ}, and the momentum constraint~\eqref{eq:MCz} relates it to $d_f^z$, in the same way as Eqs.~ \eqref{eq:dtheta} and \eqref{eq:df} do for the even-parity sector. 
When the background gauge field and shear are absent from the outset (the scalar-only case), the metric odd-parity perturbations reduce to a constant plus the standard decaying tensor solution proportional to $I_\alpha$, while the gauge-field perturbation decouples at linear order. 
This should be distinguished from taking $d_A\to0$ at fixed $d^z_\perp/d_A$ in expressions derived for a nonzero electric background.
In the anisotropic background, by contrast, the $c_E^z$ mode continues to evolve through the factor $e^{-6\beta}$ and sources $\delta A_z \propto c_E^z \bar{A}_{x}$ through the background electric field.

As in the even-parity sector discussed in Sec.~\ref{SS}, the odd-parity solutions can be interpreted in terms of the background, although with an important difference.
A rotation of the anisotropy axis and the gauge field toward the $z$-axis by a small time-independent angle $\Theta(x,y)$ generates $E_{z}=(e^{-6\beta}-1)\Theta$ and $\delta A_{z}=\Theta\bar{A}_{x}$.
These solve Eqs.~\eqref{eq:dAsol} and \eqref{eq:Ezsol2} with $c^{z}_{E}=d^{z}_{\perp}/d_{A}=\Theta$ and $d^{z}_{E}=c^{z}_{\delta A}=0$, and Eq.~\eqref{eq:dEz} together with $He^{3\alpha}\beta'=d_{\beta}+d_{A}\bar{A}_{x}/3$ gives $\tilde{d}^{z}_{E}=-6d_{\beta}\Theta$.
For a constant $\Theta$, the rotation is a global coordinate transformation.
For a spatially varying $\Theta$, however, the momentum constraint~\eqref{eq:MCz} requires $d^{z}_{f}=-6i\,d_{\beta}(k_{x}/k^{2}_{y})\,\Theta$.
In the even-parity sector, the rotation is a combination of the residual transformations $L^{x}$ and $L$ and thus satisfies the constraints automatically.
The odd-parity sector has no counterpart of $L^{x}$, so that the rotated patch must be accompanied by the decaying mode of $f_{z}$ whenever the background shear $d_{\beta}$ is nonzero.
The expression for $d_f^z$ assumes $k_y\neq0$; it is not a formula for the exactly parallel direction.
The constants $d^{z}_{E}$ and $d^{z}_{f}$ correspond to the variations of the off-diagonal shear charges of a general Bianchi type I background in the $(x,z)$ and $(y,z)$ directions, respectively.

\section{Non-linear perturbations}
\label{NL}

In this appendix, we examine a nonlinear extension of the separate-universe picture at leading order in the spatial gradient expansion, $O(\epsilon^0)$. 
We restrict the shear evolution to the sector aligned with the conserved electric-flux projector, neglecting non-aligned decaying shear components after a reference time $\tau_0$. 
An aligned homogeneous shear component can still be included.

At leading order in the gradient expansion, the separate-universe evolution is guaranteed on general grounds~\cite{Tanaka:2021dww}.
However, in the presence of a large-scale anisotropic stress, the spatial metric is in general given only formally.
The shear is sourced by the anisotropic stress, which itself depends on the spatial metric, $\pi_{ij}=\pi_{ij}[\gamma]$.
The solution for $\gamma_{ij}$ therefore takes the form of nested time integrals in which $\gamma_{ij}$ appears on both sides~\cite{Tanaka:2021dww}.
Within the aligned-shear sector specified below, the nested time integrals can be resolved explicitly for the present model.
The key observation is that the conserved electric flux $\mathcal{C}^{i}$ remains an eigenvector of the shear once the decaying shear mode is neglected.
As a consequence, the shear is proportional to a time-independent projection operator, and the tensorial evolution equation reduces to a single scalar equation for a function $b(\tau)$, which is identical to the background equation for $\beta$ [Eq.~\eqref{eq:CoresNL}].
The full spatial metric is then obtained in closed form, Eq.~\eqref{eq:gammasol2}, in terms of its initial value and $b(\tau)$ alone.
Within this sector the construction allows arbitrary functions $U(\phi)$ and $I(\phi)$, and does not require the local electric-flux direction to be close to the background axis. In particular, finite off-diagonal components of the spatial metric are retained non-perturbatively, rather than being assumed small from the outset~\cite{Abolhasani:2013zya,Talebian-Ashkezari:2016llx}.
It thus provides the non-linear counterpart of the background--perturbation correspondence established at linear order in Sec.~\ref{Cor}.

\subsection{Metric and basic equations}

In this subsection, we introduce the ADM decomposition and define the trace and trace-free parts of the spatial metric and extrinsic curvature. 
We also write down the components of the energy-momentum tensor for the scalar and gauge fields in this decomposition. 
Let us write the metric in the form
\begin{align}
    ds^2=-\Lps^2dt^2+\hat{\gamma}_{ij}dx^{i}dx^{j},
    \end{align}
where $\Lps$ is the lapse function.
We have chosen the threading so that the shift vector vanishes.
In this appendix, $\Lps$, $\Sh_{ij}$ and $\QA^{i}_{\ j}$ denote the lapse, the trace-free part of the extrinsic curvature (shear) and the constant matrix of the decaying shear mode, respectively; they are named so as not to be confused with the variables $L$, $A$, $C$ and $E$ of the main text.
The matter quantities $\rho$, $P$, $J_{i}$ and $\pi_{ij}$ are defined exactly as in Sec.~\ref{BG}, with $n^{\mu}$ the unit normal to the constant-time hypersurface, but are now evaluated for the full nonlinear fields; they reduce to their background values in the homogeneous limit.
Likewise, $K$ in this appendix denotes the trace of the extrinsic curvature and $V$ the local counterpart of $\mathcal{V}$.
The extrinsic curvature of the constant time hypersurface can be defined as
\begin{align}
    K_{ij}:=\frac{1}{2}\partial_{\tau}\hat{\gamma}_{ij},
    \end{align}
where $d\tau=\Lps dt$ is the proper time.
We may decompose $\hat{\gamma}_{ij}$ and $K_{ij}$ into the trace and the trace-free part as
\begin{align}
    &\hat{\gamma}_{ij}=e^{2\alpha(t)}e^{2\psi(\vec{x},t)}\gamma_{ij}(\vec{x},t), \quad \det \gamma_{ij}=1, \\
    &K_{ij}=e^{2\alpha(t)}e^{2\psi(\vec{x},t)}\qty(\frac{\gamma_{ij}}{3}K+\Sh_{ij}), \quad \gamma^{ij}\Sh_{ij}=0,
    \end{align}
where $K$ and $\Sh_{ij}$ can be written as
\begin{align}
    &K=3\partial_{\tau}(\alpha+\psi), \label{eq:KNL} \\
    &\Sh_{ij}=\frac{1}{2}\partial_{\tau}\gamma_{ij}, \label{eq:ANL}
    \end{align}
and indices of $\Sh_{ij}$ are raised and lowered by $\gamma_{ij}$.

\subsection{Field equations at leading order}

Here we present the Einstein equations in the ADM formalism. 
At leading order in the gradient expansion, the constraint equations and the evolution equations reduce to the forms given below, which will serve as the starting point for solving the nonlinear system.

In the ADM decomposition, the equations are classified by the constraint equations and the evolution equations.
In the leading order of the gradient expansion, the constraints are written as
\begin{align}
    &\frac{2}{3}K^2-\Sh_{ij}\Sh^{ij}=2\rho, \label{eq:NLHam} \\
    &\frac{2}{3}\partial_{i}K-e^{-3\psi}D_{j}(e^{3\psi}\Sh^{j}_{\ i})=J_{i}, \label{eq:NLMon} \\
    & \rho:=T_{\mu\nu}n^{\mu}n^{\nu} \quad J_{i}:=-T_{i\mu}n^{\mu}, 
    \end{align}
where $n^{\mu}:=\qty(1/\Lps,\vec{0})$ is a normal vector to the constant time hypersurface and $D_{j}$ denotes the covariant derivative with respect to $\gamma_{ij}$.
The dynamical equations are given by
\begin{align}
    &\partial_{\tau}K=-\frac{1}{2}K^2-\frac{3}{4}\Sh_{ij}\Sh^{ij}-\frac{3}{2}P, \label{eq:NLK} \\
    &\partial_{\tau}\Sh_{ij}=-K\Sh_{ij}+2\Sh_{i}^{\ k}\Sh_{kj}+\pi_{ij}, \label{eq:NLA}
    \end{align}
where the pressure terms are defined by
\begin{align}
    &T_{ij}=e^{2\alpha}e^{2\psi}\qty(P\gamma_{ij}+\pi_{ij}),   \quad \gamma^{ij}\pi_{ij}=0.
    \end{align}

As we are focusing on the case where the matter fields are given as a scalar field and U(1) gauge field, components of the energy-momentum tensor are written as 
\begin{align}   
&\rho=\frac{1}{2}(\partial_{\tau}\phi)^2+U+\frac{I^2}{2e^{2\alpha}e^{2\psi}}(\partial_{\tau}A)^2, \\
&P=\frac{1}{2}(\partial_{\tau}\phi)^2-U+\frac{I^2}{6e^{2\alpha}e^{2\psi}}(\partial_{\tau}A)^2, \\
&J_{i}=-\partial_{\tau}\phi\partial_{i}\phi+\frac{I^2}{e^{2\alpha}e^{2\psi}}\gamma^{jk}\partial_{\tau}A_{k}(\partial_{j}A_{i}-\partial_{i}A_{j}), \\
&\pi_{ij}=-\frac{I^2}{e^{2\alpha}e^{2\psi}}\qty[\partial_{\tau}A_{i}\partial_{\tau}A_{j}-\frac{\gamma_{ij}}{3}(\partial_{\tau}A)^2], \label{eq:NLPi}
    \end{align}
respectively.
Here, we have defined  $(\partial_{\tau}A)^2:=\gamma^{ij}\partial_{\tau}A_{i}\partial_{\tau}A_{j}$.
Equations of motion for the matter fields are given as
\begin{align}   
&0=-\frac{1}{\sqrt{-g}}\partial_{\nu}\qty(\sqrt{-g}g^{\mu\nu}\partial_{\mu}\phi)+U_{\phi}+\frac{1}{2}II_{\phi}F_{\mu\nu}F^{\mu\nu}, \label{eq:NLPhi} \\
&0=\frac{1}{\sqrt{-g}}\partial_{\nu}\qty(\sqrt{-g}I^2F^{\mu\nu}). \label{eq:NLMax}
    \end{align}

\subsection{Solving equations}

In this subsection, we obtain a formal integral representation of the shear, including a time-independent integration matrix $\QA^{i}_{\ j}(\mathbf x)$. 
Because the source projector depends on the spatial metric before imposing the aligned-shear restriction, this representation is generally implicit. 
It is the starting point for the restricted nonlinear construction in Sec.~\ref{C4}.

First, we shall solve the Maxwell equation~\eqref{eq:NLMax}.
From the spatial components ($\mu=i$), neglecting the spatial gradients, we obtain
\begin{align}   
&0=\partial_{t}\qty(e^{\alpha}e^{\psi}I^2\gamma^{ij}\partial_{\tau}A_{j}), \nonumber \\
\therefore &\partial_{\tau}A_{i}=\frac{\gamma_{ij}}{e^{\alpha}e^{\psi}I^2}\mathcal{C}^{j},
    \end{align}
where $\mathcal C^j(\mathbf x)$ is constant in proper time at this order, but may vary from one long-wavelength patch to another.
From the time component ($\mu=t$), i.e., Gauss's law, we can see that
\begin{align}   
&0=\partial_{i}\qty(e^{\alpha}e^{\psi}I^2\gamma^{ij}\partial_{\tau}A_{j}), \nonumber \\
\therefore &\partial_{j}\mathcal{C}^{j}=0.
    \end{align}
Using this solution, the anisotropic stress~\eqref{eq:NLPi} is rewritten as
\begin{align}   
\pi_{ij}&=-\frac{1}{I^2e^{4\alpha}e^{4\psi}}\qty(\mathcal{C}_{i}\mathcal{C}_{j}-\frac{\gamma_{ij}}{3}\mathcal{C}^2)=-\frac{\mathcal{C}^{k}}{e^{3\alpha}e^{3\psi}}\qty(\gamma_{ik}\partial_{\tau}A_{j}-\frac{\gamma_{ij}}{3}\partial_{\tau}A_{k}), \label{eq:piac}
    \end{align}
where $\mathcal{C}_{i}:=\gamma_{ij}\mathcal{C}^{j}$ and $\mathcal{C}^2:=\gamma_{ij}\mathcal{C}^{i}\mathcal{C}^{j}$.
We  note that $\mathcal{C}^{j}$ is an eigenvector of $\pi^{i}_{\ j}$.
That is to say,
\begin{align}   
\pi^{i}_{\ j}\mathcal{C}^{j}&=-\frac{2}{3}V\mathcal{C}^{i}, \quad V:=\frac{\mathcal{C}^2}{I^2e^{4\alpha}e^{4\psi}}.
    \end{align}

Raising an index in Eq.~\eqref{eq:piac}, the anisotropic stress can be written directly as
\begin{align}
&\pi^{i}_{\ j}=-\frac{V}{3}P^{i}_{\ j}, \label{eq:piP} \\
&P^{i}_{\ j}:=3P^{i}_{C, j}-\delta^i_j, \quad P^{i}_{C, j}:=\frac{\mathcal{C}^{i}\mathcal{C}_{j}}{\mathcal{C}^{2}}.
\end{align}
Here $P^{i}_{C,j}$ is the projector onto $\mathcal{C}^{i}$, so that $P^{i}_{\ j}$ has the eigenvalue $2$ on $\mathcal{C}^{i}$ and $-1$ on the two-dimensional subspace orthogonal to it, and $\delta^{i}_{\ j}-P^{i}_{C,j}$ is the projector onto this subspace.

Integrating Eq.~\eqref{eq:NLA}, whose mixed-index form reads $\partial_{\tau}\Sh^{i}_{\ j}=-K\Sh^{i}_{\ j}+\pi^{i}_{\ j}$, and using Eq.~\eqref{eq:piP}, the shear tensor can be expressed as
\begin{align}
    \Sh^{i}_{\ j}=e^{-3(\alpha+\psi)}\qty(\QA^{i}_{\ j}-\int d\tau e^{3(\alpha+\psi)}\frac{V}{3}P^{i}_{\ j} ).
    \label{eq:AijDec}
   \end{align}
The integration matrix $\QA^i_{\ j}$ is time independent (but can depend on position), and its contribution to the mixed-index shear is weighted by $e^{-3(\alpha+\psi)}$. The associated contribution to the spatial metric involves an additional time integral, so its size must be assessed for the background under consideration.

\subsection{Growing-mode solution of the spatial metric}
\label{C4}

The shear representation~\eqref{eq:AijDec} is now restricted by dropping, for $\tau\geq\tau_0$, the components of $\QA^i_{\ j}$ not aligned with $P^i_{\ j}$. 
The component proportional to $P^i_{\ j}$ can be retained and absorbed into the scalar coefficient $\Sigma$, just as the homogeneous shear integration constant $d_\beta$ is retained in the background. 
This is a consistent aligned-shear sector, not the full nonlinear solution with arbitrary decaying shear. For expanding patches the contribution of non-aligned decaying shear to $\gamma_{ij}$ may approach a constant, and the offset accumulated before $\tau_0$ can be included in $\gamma_{ij}(\tau_0)$. 
However, the omitted evolution after $\tau_0$ is not exactly encoded by that initial metric: it is negligible only when the remaining time integral of the decaying shear is sufficiently small. We therefore do not assume a universal suppression factor independent of the background or the shear history.
At linear order, $\QA^{i}_{\ j}$ corresponds to the Noether charges $d_{E}$ and $d_{f}$ of Sec.~\ref{SS}, which are fixed by the momentum constraints~\eqref{eq:dtheta} and \eqref{eq:df} in terms of the gradients of $(W,\delta d_{\beta},\delta d_{A})$. 
In particular, they need not vanish even when $W=\delta d_{\beta}=0$, if $\delta d_{A}$ has a gradient.
The closed-form metric below describes the subsequent evolution within this restricted sector, with off-diagonal metric data already present at $\tau_0$ retained in $\gamma_{ij}(\tau_0)$. 
It does not include any further contribution from the omitted non-aligned decaying shear.
The constraints, i.e., the momentum constraint~\eqref{eq:NLMon} and the Gauss law $\partial_{j}\mathcal{C}^{j}=0$, are to be imposed on these initial data. 
As pointed out in Ref.~\cite{Tanaka:2021dww}, they are not automatically satisfied when the gauge field does not decay on large scales.

The analysis consists of two parts.
First, we reduce the tensorial shear equation to a scalar eigenvalue equation and compare it with the background in Sec.~\ref{Red}.
Then, we explicitly integrate the spatial metric and read off the correspondence with the linear perturbation variables in Sec.~\ref{Integgamma}.

\subsubsection{Reduction to scalar equation and comparison with background}
\label{Red}

Within the aligned-shear sector, we show that the shear tensor is proportional to a time-independent projection operator. 
This reduces the tensorial evolution equation to a single scalar equation for $\Sigma$.
By choosing a gauge aligned with the gauge-field momentum, we demonstrate that this scalar equation is algebraically identical to the background equation for the anisotropy parameter $\beta$, thereby establishing the nonlinear separate-universe correspondence at leading order in the gradient expansion for the growing modes.

When the non-aligned part of $\QA^{i}_{\ j}$ is omitted, we have $\QA^{i}_{\ j}=q\,P^{i}_{\ j}$ with a time-independent coefficient $q$.
At each time, $P^i_{\ j}\mathcal C^j=2\mathcal C^i$, while $\mathcal C^i$ itself is time independent.
Both terms in~\eqref{eq:AijDec} therefore have $\mathcal C^i$ as an eigenvector, and the shear satisfies
\begin{align}
    &\Sh^{i}_{\ j}\mathcal{C}^j = 2\,e^{-3(\alpha+\psi)}\qty(q-\frac{1}{3}\int d\tau\, e^{3(\alpha+\psi)}V)\mathcal{C}^{i} =: 2\Sigma\,\mathcal{C}^{i}.
    \end{align}
The constant $q$ is the counterpart of $d_{\beta}$ in Eq.~\eqref{eq:Acbeta}; it enters the initial shear $\partial_{\tau}b(\tau_{0})$ of each patch and is not dropped in what follows.
The function $\Sigma$ is determined as follows.
By taking the time-derivative of $\mathcal{C}^2$, we have
\begin{align} 
&\partial_{\tau}\mathcal{C}^{2} = 2\Sh_{ij}\mathcal{C}^{i}\mathcal{C}^{j} = 4  \Sigma \mathcal{C}^{2}. \\
&\therefore \Sigma = \frac{\partial_{\tau}\mathcal{C}^{2}}{4 \mathcal{C}^{2}}. \label{eq:X}
\end{align} 
Furthermore, we can show that $P^{i}_{ C,j}$ is independent of the time $\tau$:
\begin{align} 
&\partial_{\tau}\mathcal{C}_{i} = 2\Sh_{ij}\mathcal{C}^{j} = 4 \Sigma\mathcal{C}_{i}, \\
\therefore  & \partial_{\tau} P^{i}_{C, j} = \frac{\mathcal{C}^{2}\partial_{\tau}\qty(\mathcal{C}^{i}\mathcal{C}_{j}) -\partial_{\tau} \mathcal{C}^{2} \cdot \mathcal{C}^{i}\mathcal{C}_{j}}{\qty(\mathcal{C}^{2})^2} =0.
\end{align}  
Therefore, we can write the shear as
\begin{align}   
&\Sh^{i}_{\ j}=\Sigma P^{i}_{\ j}. \label{eq:AX}
    \end{align}  
By substituting this relation into the mixed-index form of Eq.~\eqref{eq:NLA}, we have
\begin{align}
    &\partial_{\tau} \Sigma+K \Sigma=-\frac{1}{3}V \label{eq:NLAC}.
    \end{align}

The complementary projector $\delta^i_{\ j}-P^i_{C,j}$ is likewise independent of proper time.
The term proportional to $q$ in $\Sigma$ is the homogeneous solution of this equation.

Locally in a region with $\mathcal C^i\neq0$, choose spatial coordinates adapted to the electric-flux vector so that $\mathcal C^i=\mathcal C^x\delta^i_x$.
We also write  $\gamma_{xx} = e^{-4b}\Gamma_{xx}$, where $\Gamma_{xx}$ is independent of $\tau$.
Then, from Eq.~\eqref{eq:X}, we obtain
\begin{align}   
&\Sigma = -\partial_{\tau}b,
    \end{align}
and Eqs.~\eqref{eq:NLHam}, \eqref{eq:NLK} and Eq.~\eqref{eq:NLAC} can be written as
\begin{align}
&-\frac{1}{3}K^2+3 \qty(\partial_{\tau}b)^2=-\rho, \\
    &\frac{2}{3}\partial_{\tau}K+\frac{K^2}{3}+3 \qty(\partial_{\tau}b)^2=-P, \label{eq:NLKC} \\
    &\partial^2_{\tau} b +K \partial_{\tau}b =\frac{1}{3}\frac{\Gamma_{xx}\qty(\mathcal{C}^x)^2}{I^2e^{4\alpha}e^{4\psi}e^{4b}} \label{eq:NLACb}.
    \end{align}
By comparing these EoMs with the background equations~\eqref{eq:ABG00}, \eqref{eq:ABGtr}, \eqref{eq:Abeta1eq} with $\gamma=0$, one can see that they are equivalent under the identifications
\begin{align} 
\tau \leftrightarrow t, \quad
 K\leftrightarrow 3H, \quad   b \leftrightarrow \beta, \quad e^{\alpha+\psi}\leftrightarrow e^{\alpha}, \quad \Gamma_{xx}(\mathcal{C}^{x})^{2}\leftrightarrow d^{2}_{A}, \label{eq:CoresNL}
    \end{align}
up to the momentum constraint, which is absent in the background.
Thus, within the aligned-shear sector and subject to the local constraints, the nonlinear superhorizon evolution is governed by the same equations as a homogeneous Bianchi I background.
As we will see in the next subsubsection, once $b(\tau)$ is obtained, the full spatial metric $\gamma_{ij}(\tau)$ is determined in terms of the initial data $\gamma_{ij}(\tau_0)$.
To apply the above correspondence in practice, one first picks a reference background solution $(\beta, \phi, \bar{A}_{x})$.
For each patch in this sector, the evolution of $b(\tau)$ follows from the corresponding background equations with its local initial scalar, flux, expansion and shear data; no linearization in the initial anisotropy or flux direction is required.

\subsubsection{Integration of spatial metric and correspondence with linear perturbations}
\label{Integgamma}

Having established the correspondence in Sec.~\ref{Red}, we now explicitly solve Eq.~\eqref{eq:AX} for the spatial metric $\gamma_{ij}$. 
We can express the full metric in terms of the initial data and the function $b(\tau)$, and then read off the correspondence between the nonlinear variables and the linear perturbation variables $(D, E, f)$ introduced in Sec.~\ref{SPG}.

From Eqs.~\eqref{eq:ANL} and \eqref{eq:AX}, we obtain 
\begin{align}   
&\gamma_{i j}(\tau)=\gamma_{i k}(\tau_0)\exp\qty(2\mathcal{X}P)^k_{\, j}, \label{eq:gammasol} \\ 
&\mathcal{X}(\tau) := \int_{\tau_0}^{\tau}  \Sigma(\tau_{*} ) d\tau_{*}.
    \end{align}
Since $P^i_{\ j}$ is traceless, the matrix exponential has unit determinant, consistent with $\det\gamma=1$.
Furthermore, we can write
\begin{align}   
&\exp\qty(2\mathcal{X} P)^i_{\, j} =e^{-2\mathcal{X}} \delta^i_j + \qty(e^{4\mathcal{X}}-e^{-2\mathcal{X}}) P^i_{C, j}, \\   
&\therefore \gamma_{i j}(\tau)=
e^{-2\mathcal{X}}\gamma_{i k}(\tau_0) + \qty(
e^{4\mathcal{X}}-e^{-2\mathcal{X}})\gamma_{i k}(\tau_0)P^k_{C, j}. \label{eq:gammasol2}
    \end{align}

Equation~\eqref{eq:gammasol} is obtained without specifying the form of $\mathcal{C}^i$.
In the local flux-adapted coordinates introduced above, where $\mathcal C^i=\mathcal C^x\delta_x^i$ and $\Sigma=-\partial_\tau b$, we have
\begin{align}   
&\mathcal{X}(\tau) = -b(\tau) +b(\tau_0), \\
&P^i_{C, j} = \delta^{i}_{x}\qty(\delta^x_j + \frac{\gamma_{xa}}{e^{-4b}\Gamma_{xx}}\delta^a_j).
    \end{align}
Substituting these relations into Eq.~\eqref{eq:gammasol}, we obtain
\begin{align}   
&\gamma_{xx}(\tau)=
e^{4\mathcal{X}}\gamma_{xx}(\tau_0)=e^{-4b(\tau)}\Gamma_{xx}, \quad \Gamma_{xx} =  e^{4b(\tau_0)}\gamma_{xx}(\tau_0), \label{eq:gammaxx} \\
&\gamma_{xa}(\tau)=
e^{4\mathcal{X}}\gamma_{xa}(\tau_0), \label{eq:gammaxa} \\
&\gamma_{ab}(\tau)=
e^{-2\mathcal{X}}\gamma_{ab}(\tau_0) + \qty(
e^{4\mathcal{X}}-e^{-2\mathcal{X}})\frac{\gamma_{xa}(\tau_0)\gamma_{xb}(\tau_0)}{\gamma_{xx}(\tau_0)}. \label{eq:gammaab}
    \end{align}

Therefore, the components of the spatial metric $\gamma_{ij}$ can be solved once their initial values and  $b(\tau)$ are specified.
As we have discussed above, $b(\tau)$ can be obtained by varying the background solutions for $\beta$.

It is instructive to clarify the correspondence between 
the nonlinear solutions \eqref{eq:gammaxx}--\eqref{eq:gammaab} and the linear perturbation variables 
$D, E, f$ introduced in Sec.~\ref{SPG}.
Linearizing about the axisymmetric reference background, with the same spatial normalization and residual-gauge convention as in Sec.~\ref{SPG}, gives
\begin{align}   
&D = \frac{1-e^{4\beta}\gamma_{xx}}{4} \simeq b-\beta, \quad \partial_a E = e^{-2\beta}\gamma_{xa}, \quad f_{ab} = \frac{e^{-2\beta} }{2}\qty(\gamma_{ab}- \frac{\delta_{ab}}{2}\delta^{cd}\gamma_{cd}).
    \end{align}
At linear order, the initial off-diagonal and traceless metric data contain residual spatial-gauge contributions (identified with $c_E$ and $c_f$ in Sec.~\ref{SS}) as well as physical contributions. The latter may include shear memory accumulated before $\tau_0$, represented by the integration constants $d_E$ and $d_f$ in the full linear solutions~\eqref{eq:Esol2} and \eqref{eq:Fsol}. Equations~\eqref{eq:gammaxa} and \eqref{eq:gammaab} transport these initial data, but omit the additional evolution sourced by the non-aligned decaying-shear components for $\tau>\tau_0$. 
The resulting closed-form metric demonstrates the nonlinear counterpart of the background--perturbation correspondence in this aligned-shear sector at leading order in the spatial gradient expansion.
It is not a reconstruction of the full linear memory terms associated with the discarded non-aligned shear charges after $\tau_0$.

\bibliography{bibs/hoge}
\bibliographystyle{unsrt}

\end{document}